\documentclass[11 pt]{article}
\usepackage[utf8]{inputenc}
\usepackage[english]{babel}
\usepackage{amsmath}
\usepackage{enumerate}
\usepackage{amsthm}
\usepackage{latexsym,ulem}
\usepackage{amssymb}
\usepackage[pdftex]{graphicx}
\usepackage{float}
\usepackage{natbib}
\let\natbibcitet\citet
\renewcommand\citet{\bibpunct{(}{)}{,}{a}{,}{,}\natbibcitet}
\let\natbibcitep\citep
\renewcommand\citep{\bibpunct{(}{)}{;}{a}{,}{;}\natbibcitep}
\usepackage{textcomp}
\usepackage{appendix}
\usepackage{bibentry}
\usepackage{xcolor}
\theoremstyle{definition}

\newtheorem{Prop}{Proposition}

\newtheorem{Rem}{\textit{Remark}}

\newcommand{\argmax}[2]{%
\smash{\mathop{{\rm argmax}}\limits_{#1}}\,#2} 
\def\ee{{\mathbb E}}

\def\pp{{\mathbb P}}
\def\vv{{\mathbb V}}
\def\rr{{\mathbb R}}

\newcommand{\mb}[1]{\mathbf{#1}}
\newcommand{\ms}[1]{\boldsymbol{#1}}
\newcommand{\dd}{\mathrm{d}}

\newcommand{\NNew}[1]{\textcolor{black}{#1}}
\newcommand{\NNNew}[1]{\textcolor{black}{#1}}

\usepackage[textwidth=16cm,footskip=1.5cm,textheight=23cm,marginratio=1:1,heightrounded]{geometry}
\author{Guillaume Damblin\\
EDF R\&D MRI, Chatou, France\\
UMR MIA-Paris,\\ AgroParisTech, INRA,\\ Université Paris-Saclay,\\ 75005, Paris, France
\and 
Pierre Barbillon\footnote{Corresponding author: 16 rue Claude Bernard, 75005 Paris}
\footnote{\textsc{pierre.barbillon@agroparistech.fr}} \\
UMR MIA-Paris,\\ AgroParisTech, INRA,\\ Université Paris-Saclay,\\ 75005, Paris, France
\and 
Merlin Keller\\
EDF R\&D MRI, Chatou, France
\and 
Alberto Pasanisi\\
EDF R\&D - EIFER, Karlsruhe, Germany
\and
\'Eric Parent\\
UMR MIA-Paris,\\ AgroParisTech, INRA,\\ Université Paris-Saclay,\\ 75005, Paris, France
}
\title{Adaptive numerical designs for the calibration of computer codes}

\begin{document}

\maketitle

\begin{abstract}
Making good predictions of a physical system using a computer code requires the inputs 
to be carefully specified. Some of these inputs, called control variables, reproduce physical conditions whereas other inputs, called parameters, are specific to the computer code and most often uncertain. The goal of statistical calibration consists in reducing their uncertainty with the help of a statistical model which links the code outputs with the field 
measurements. In a Bayesian setting, the posterior distribution of these parameters is typically sampled using MCMC methods. However, they are impractical when the code runs are highly time-consuming. A way to circumvent this
issue consists of replacing the computer code with a Gaussian process emulator, then sampling a surrogate posterior distribution based on it. Doing so, calibration is subject to an error which strongly depends
on the numerical design of experiments used to fit the emulator.
{Under the assumption that there is no code discrepancy,} we aim to reduce this error by constructing a sequential design by means of the Expected Improvement criterion.
Numerical illustrations in several dimensions assess the efficiency of such sequential strategies.


\end{abstract}

\section{Introduction}

This work is incorporated within the
field of uncertainty quantification in computer experiments. 
A crucial issue in engineering (aerospace, car, nuclear, etc.) concerns the ability of 
computer codes (also called simulators or computer models) to mimic a physical phenomenon of interest as well as possible. 
In this regard, the field of so-called Verification and Validation (V$\&$V)  aims to assess 
the accuracy of computer predictions for many applications. For instance, the study of V$\&$V has become a huge preoccupation in 
the nuclear industry where numerical simulation is more and more used to assess the safety of installations for which physical experiments are 
impractical or economically infeasible. An essential prerequisite of V$\&$V consists in quantifying all sources of uncertainty involved in a code output \citep{Roy11}. 
Our paper is focused on the reduction of one of them, called parameter uncertainty, caused by the lack of knowledge about the value of parameters which are specific to the computer code \citep{Koh2001}. They can be either non-measurable
physical quantities or just tuning factors. Calibration comes down to a statistical inference of
these parameters after assuming a statistical model which makes explicit the relationship between the code outputs and the field measurements \citep{Cam06,Cox01}. Another popular framework which deals with parameter uncertainty is called History Matching (HM) \citep{Craig97}. HM can detect regions in the parameter space which appear to be incompatible with the field measurements. Based on an implausibility measure, 
this method is well-suited for large systems wherein the size of inputs makes immediate calibration intractable.   
In the same way, Sensitivity Analysis (SA) can detect which parameters have a negligible effect on the code output \citep{Salt00}. 
Hence, both HM and SA can shrink the input space before carrying out calibration. 

Our work focuses on calibration in a Bayesian fashion \citep{Koh2001,Bay07}, rather than a frequentist one \citep{Lop06,Gramacy14,Wong14}.
This strategy of inference provides an appropriate framework to quantify the parameter uncertainty from prior
to posterior distribution as new data become available. 
In the literature, Bayesian calibration is usually performed within a framework where the code predictions 
suffer from a systematic discrepancy for any value of parameters, which reflects the view that the mathematical equations underlying the code should not be considered as a perfect model of the real world \citep{Koh2001,Higd04}. Even if this framework is more realistic, some confounding can appear when the parameters and the shape of the discrepancy are jointly estimated \citep{Lop06,Bry_OHagan14}. 
Because this issue is outside the scope of this paper, our presentation is centered on a statistical model which does not include the code discrepancy \citep{Cox01}.
However, it would be possible to generalize our framework if the shape of the discrepancy were provided by prior expertise. 

We deal with Bayesian calibration when the code runs are time-consuming, a critical issue frequently arising in the field of computer experiments. Indeed, when a simulation needs several hours, even several days to run, then the MCMC algorithms become unfeasible.
For instance, if each simulation lasts one hour, then $10, 000$ simulations launched by an MCMC exploration of the parameter space will require more than a year,
making the calibration process impractical.
A well-known solution to this issue is to replace the code in the likelihood expression by a Gaussian process emulator (GPE),
constructed from a training set of simulations over a set of input locations, the so-called design of (numerical) experiments.
In this paper, we provide a theoretical result validating this approach under quite standard hypotheses.
Then, we propose two new algorithms for constructing sequential designs aiming at reducing the calibration error 
induced by the uncertainty of the GPE when the number of possible simulations is fixed.
Although it was already mentioned by \citet{Koh2001} as an important research axis, few papers deal with that issue. To the best of our knowledge, \citet{Kumar} has already proposed some 
empirical criteria for sequentially selecting the code runs.
\citet{Prat13} have also proposed adaptive strategies whereby the Expected Improvement (EI) criterion is computed over a likelihood ratio. {The EI criterion is much used for construction of designs of experiments} in order to optimize black box functions when only a limited number of simulations can be run \citep{Jones98}.
Recently, it has been applied to solve a problem of optimization under uncertainty when the code 
inputs are $(\mb{x},\mb{u})$, where $\mb{x}$ denotes a vector of control variables and $\mb{u}$ denotes a vector
of random variables \citep{Janis12}. In the same spirit, we propose new algorithms which consist in applying the EI criterion to the sum of squares of the residuals between the code outputs and the field measurements when the code inputs are $(\mb{x},\ms{\tau})$ where $\ms{\tau}$ is a vector of parameters. 
In this way, we aim at reducing the uncertainty due to the GPE in actual regions of high posterior density.
We emphasize that, contrary to \citet{Prat13}, we propose to emulate the computer code instead of emulating a 
function of the code. This will permit to update the EI criterion from a single evaluation of the code only.
Numerical simulations show that these sequential designs lead to make the surrogate posterior distribution (SPD) of
the parameters (where the code is replaced with the GPE)
closer to the actual posterior distribution (APD).
\NNNew{Contrary to \citet{conrad2016} who introduced a Metropolis-Hastings algorithm with local approximations of the code where these approximations
are refined along the chain with new runs, we first construct the adaptive design based on EI criterion.
From this design, 
a global approximation of the code as a GPE is computed and then the calibration is conducted as in \citet{Koh2001} or in \citet{Higd04}}.

This paper is divided into five sections.
In Section \ref{sec:two_calib}, the statistical framework is introduced and the main features of the
GPE are recalled. In Section \ref{sec:three_calib}, two new algorithms for Bayesian calibration based on 
the Expected Improvement criterion are laid out. Their performances are illustrated on two academic examples in Section \ref{sec:four_calib}.
The conclusions of this paper are provided in Section \ref{sec:five_calib}. 


\section{Calibration framework}

\label{sec:two_calib}

\subsection{Notations and modeling}
Let $r(\mb{x})\in\rr$ be a physical quantity of interest
where $$\mb{x}=\big(x_1,\cdots,x_d\big)^{T}\in\mathcal{X}\subset\rr^{d}$$ is a vector of control variables. This kind of variable is measurable in the field experiments
and characterizes the system. It can include both physical variables (temperature, pressure, velocity, etc.) 
and design variables (height, area, etc.). We suppose 
that a number of field experiments, say $n$, has been collected.
In this paper, the locations of field experiments will be referred to as the matrix $$\mb{X}^{f}=\big[\mb{x}^f_1,\cdots,\mb{x}^f_n\big]^{T} \in M_{n,d}(\rr)$$ and the corresponding measurements will be referred to as the vector $\mb{z}^f=(z^f_1,\cdots,z^f_n)^{T}$. Due to observation errors, $\mb{z}^f$ is not exactly equal to $r(\mb{X}^f)=(r(\mb{x}^f_1),\cdots,r(\mb{x}^f_n))$. Hence, for $1\leq i\leq n$,

\begin{equation}
\label{eq2}
z^f_i=r(\mb{x}^f_i)+\epsilon_i,
\end{equation}
where
\begin{equation*} 
{\epsilon_i}\underset{i.i.d.}{\thicksim}
\mathcal{N}(0,\lambda^{2})\,,
\end{equation*}
is modeled as a white noise. The variance $\lambda^2$ is assumed to be known 
for the sake of simplicity. 
{In many cases, the nature of the observation errors is sufficiently well characterized that their distribution can be treated as known. For example, in the case where the measurement precision is the only source of observation error at stake, the precision of the measuring device is either known or can be estimated from replicates.} 

Let $y_{\ms{\tau}}(\mb{x})$ be a deterministic computer code which predicts $r(\mb{x})$ where $\ms{\tau}=(\tau_1,\cdots,\tau_p)^{T}\in\mathcal{T}\subset\rr^{p}$ is a vector of parameters including either factors attached to the field (chemical rate, friction coefficient, etc.) or
mathematical tuning factors such as a discretization step having no counterpart in physics,
or perhaps both \citep{Gang09}. The computer code is seen as a black box function, which supposes nothing is known about the connection between the inputs $(\mb{x},\ms{\tau})$ and the code output $y_{\ms{\tau}}(\mb{x})$, also called simulation.
A numerical design of experiments refers to a set of input locations where the code is run \citep{Owen}. According to \citet{Koh2001}, the computer code should be considered as an imperfect representation of the phenomenon $r$.
Hence,
\begin{equation}
\label{model_biais1}
r(\mb{x})=y_{\ms{\theta}}(\mb{x})+b(\mb{x}),
\end{equation} 
where $b(\mb{x})$ is the code discrepancy and $\ms{\theta}$ is the true value of parameters in a certain sense. Combining (\ref{model_biais1}) and (\ref{eq2}), the statistical model which links the simulations to the field measurements is written as
\begin{equation}
\label{model_biais}
{z^f_i}=y_{\ms{\theta}}(\mb{x}^f_i)+b(\mb{x}^f_i)+{\epsilon_i}.
\end{equation} 
{The estimation of $\ms{\theta}$ in Equation (\ref{model_biais}) requires to specify a prior distribution on $b(\mb{x})$.}
This issue has been widely studied over the past decade. \citet{Koh2001,Higd04,Bay07,Liu09,Bry_OHagan14} and many others have modeled $b(\mb{x})$ by a Gaussian process. \citet{Joseph09} have proposed to use a more common regression model instead. 

Some authors question whether $b(\mb{x})$ should systematically be taken into account and have proposed strategies of model selection \citep{Lop06,Dam16}
\NNew{to decide between a model without discrepancy and a model with discrepancy. }
From now on, \NNew{ we assume that $b(\mb{x})$ cannot be distinguished from the white noise error process assumed for $\epsilon_i$. }
If not, provided that $b(\mb{x})$ is elicited from prior expertise, the method developed in this paper could still be applied.

\paragraph*{The unbiased model} 
Equation \ref{model_biais} becomes:
\begin{equation}
\label{eq3}
z^f_i=y_{\ms{\theta}}(\mb{x}^f_i)+\epsilon_i.
\end{equation}
We have chosen to conduct {Bayesian estimation for $\ms{\theta}$} because it has been shown better suited than the standard MLE\footnote{Maximum Likelihood Estimation},
where flat likelihood may need regularization, especially if the dimension of $\ms{\theta}$ is high \citep{Kumar}.
{Moreover, the uncertainty on the calibrated parameters is harder to obtain
from an MLE approach than from the posterior distribution.}
Let $y_{\ms{\theta}}(\mb{X}^{f}):=\big(y_{\ms{\theta}}(\mb{x}^f_1),\cdots,y_{\ms{\theta}}(\mb{x}^f_n)\big)^{T}$ be the vector of code outputs running over the input field data $\mb{X}^{f}$.
{Let $\pi(\ms{\theta})$ denote the prior distribution on $\ms{\theta}$.}
The posterior distribution of $\ms{\theta}$ given by the Bayes formula is the normalized version of the following product:
\begin{align}
\label{eq4}
\nonumber
\pi(\ms{\theta}|\mb{z}^f)\propto &\,\,\, \mathcal{L}(\ms{\theta};\mb{z}^f)\pi(\ms{\theta}), \\
\propto &\,\,\, \frac{1}{(\lambda)^{n}}\exp\Big[{-\frac{1}{2\lambda^2}SS(\ms{\theta})\Big]}\pi(\ms{\theta}),
\end{align}
where
\begin{equation}
\label{MC}
SS(\ms{\theta})=||\mb{z}^f-y_{\ms{\theta}}(\mb{X}^{f})||^{2}
\end{equation}
is the sum of the squares of the differences between the simulations $y_{\ms{\theta}}(\mb{X}^f)$ and the field measurements, {in other words}, the sum of the squares of the residuals.
Throughout this paper, $\pi(\ms{\theta}|\mb{z}^f)$ refers to the actual posterior distribution (APD). No closed-form expression can be obtained  
for $\pi(\ms{\theta}|\mb{z}^f)$ because $y$ is usually highly non linear with respect to $\ms{\theta}$.
In such cases, $\pi(\ms{\theta}|\mb{z}^f)$ 
needs to be sampled by running an MCMC algorithm which converges to $\pi(\ms{\theta}|\mb{z}^f)$ over a very large number of samples, often several thousands \citep{Robert+98}. In our framework, the MCMC algorithms are thus unfeasible
since each sample requires a time-consuming simulation. 
A way to address this issue is to treat $y(\cdot)$ as an unknown function and setting a Gaussian process prior on it, denoted by $Y(\cdot)$,
where $(.)$ corresponds to a pair of inputs $(\mb{x},\ms{\tau})$. {In what follows, we use the notation $\ms{\tau}\in\mathcal{T}$ to
distinguish any value of the code parameter from $\ms{\tau}=\ms{\theta}$ which refers to the true value to be estimated.}



\subsection{Gaussian process emulator}
The Gaussian process was first introduced within the field of computer experiments by \citet{Sacks89}. It is the most familiar surrogate model used to mimic a costly computer code. From a Bayesian perspective, the Gaussian process, denoted in this paper by $Y(.)$, should be considered as a prior structure on the code \citep{Currin91}:
\begin{equation}
\label{GP_prior}
Y(\cdot)\sim\mathcal{GP}(m_{\ms{\beta}}(.),\sigma^{2}C_{\ms{\Psi}}(.\,,.))\,,
\end{equation}
where
\begin{itemize}
\item $m_{\ms{\beta}}(.)=h(.)^{T}\ms{\beta}$ where $h(.)=(h_1(.),\cdots,h_s(.))^{T}$ is a vector of regression functions and 
$\ms{\beta}\in\rr^{s}$ is a vector of location parameters,
\item $\sigma^{2}$ is the variance of the process,
\item $C_{\ms{\Psi}}(.\,,.)$ is the correlation function where $\ms{\Psi}$ 
is a vector of hyper-parameters including a range parameter and possibly a smoothness parameter.
\end{itemize}
{The function $C_{\ms{\Psi}}(.,.)$ encodes a prior 
information on the mathematical properties of the code output such as regularity. In some cases, this information can be available from experts of the physical phenomenon which is modeled by the code.}
Moreover, for both practical and theoretical reasons, a stationary function is almost always specified \citep{Stein99}.
For a discussion on the choice of regression and correlation functions, {refer to} \citet{koehler:owen:1996,fang:2005}.
{For $M$ runs of the computer code,} let $\mb{D}_M\in(\mathcal{X}\times\mathcal{T})^{M}$ denote the numerical design of experiments:
\begin{equation}
\label{DOE}
\mb{D}_M:=\big[(\mb{x}^D_1,\ms{\tau}^D_1),\cdots,(\mb{x}^D_M,\ms{\tau}^D_M)\big]^{T}\in \mathcal{M}_{M,d+p}(\rr),
\end{equation}
where $\mathcal{M}_{M,d+p}(\rr)$ is the space of matrices 
with $M$ rows and $d+p$ columns \NNew{with
entries in $\mathbb{R}$}.
After running the code over $\mb{D}_M$, $M$ simulations can be collected: 
\begin{equation}
y(\mb{D}_M):=\Big(y(\mb{x}^D_1,\ms{\tau}_1^D):=y_{\ms{\tau}^D_1}(\mb{x}^D_1),\cdots,y(\mb{x}^D_M,\ms{\tau}^D_M):=y_{\ms{\tau}^D_M}(\mb{x}^D_M)\Big)^{T}.
\end{equation}
Let $\mb{v}_{pred}$ and $\mb{v}_{pred}'$ be two vectors in $\mathcal{X}\times\mathcal{T}$. Then,
\begin{itemize}
\item $\Sigma_{\ms{\Psi}}(\mb{D}_M)=C_{\ms{\Psi}}((\mb{x}^D_i,\ms{\tau}^D_i),(\mb{x}^D_j,\ms{\tau}^D_j))_{1\le i,j\le M}
$ is the matrix of correlations between the simulations $y(\mb{D}_M)$,
\item $\Sigma_{\ms{\Psi}}(\mb{v}_{pred},\mb{D}_M)=(C_{\ms{\Psi}}(\mb{v}_{pred},(\mb{x}^D_i,\ms{\tau}^D_i))_{1\le i\le M}$ is the vector of correlations between $Y(\mb{v}_{pred})$ and each of $y(\mb{D}_M)$.
\end{itemize}
By conditioning the Gaussian process (\ref{GP_prior}) on the training set of simulations $y(\mb{D}_M)$, the resulting process is still a Gaussian process:
\begin{equation}
\label{eq6}
Y_{M}(.):=Y(.)|y(\mb{D}_M)\thicksim\mathcal{GP}(\mu^{M}_{\ms{\beta},\ms{\Psi}}(.),V^{M}_{\ms{\Psi},\sigma^2}(\cdot,\cdot)),
\end{equation}
with the standard expressions for the conditional mean and covariance:

\begin{align}
\label{condmean}
\nonumber
\mu^{M}_{\ms{\beta},\ms{\Psi}}(\mb{v}_{pred})=&\ee[Y_{M}(\mb{v}_{pred})]\\
=&m_{\ms{\beta}}(\mb{v}_{pred})+\Sigma_{\ms{\Psi}}(\mb{v}_{pred},\mb{D}_M)^{\textrm{T}}\Sigma_{\ms{\Psi}}(\mb{D}_M)^{-1}\Big[y(\mb{D}_M)
-m_{\ms{\beta}}(\mb{v}_{pred})\Big]\,,
\end{align}
and
\begin{align}
\label{cond_var}
\nonumber
V^{M}_{\ms{\Psi},\sigma^2}(\mb{v}_{pred},\mb{v}_{pred}')=&\,\,\textrm{Cov}(Y_M(\mb{v}_{pred}),Y_M(\mb{v}_{pred}'))
\\
=&
\,\,\sigma^{2}\Big(C_{\ms{\Psi}}(\mb{v}_{pred},\mb{v}_{pred}')-\Sigma_{\ms{\Psi}}(\mb{v}_{pred},\mb{D}_M)^{\textrm{T}}
\Sigma_{\ms{\Psi}}(\mb{D}_M)^{-1}\Sigma_{\ms{\Psi}}(\mb{v}_{pred}',\mb{D}_M)\Big).
\end{align}

\medskip
\noindent
The GPE is given by the conditional process (\ref{eq6}) which {delivers} a stochastic prediction of the code for any input $\mb{v}_{pred}$ of
the input space $\mathcal{X}\times\mathcal{T}$. In the case where $\mb{v}_{pred}$ belongs to
$\mb{D}_M$, the GPE interpolates the simulations $y(\mb{D}_M)$, that is for $1\leq i\leq M$:
\begin{equation}
 \label{eq:interpolator}
 \begin{aligned}
&  \mu^{M}_{\ms{\beta},\ms{\Psi}}(\mb{x}^D_i,\ms{\tau}^D_i)=y(\mb{x}^D_i,\ms{\tau}^D_i)\,,\\
&V^{M}_{\ms{\Psi},\sigma^2}((\mb{x}^D_i,\ms{\tau}^D_i),(\mb{x}^D_i,\ms{\tau}^D_i))=0  
 \end{aligned}
\end{equation}
which is expected for such an emulator of a deterministic computer code.
Lastly, the capability of a GPE to well predict the code should be checked against some validation criteria \citep{Bast08}. For more details about the GPE, 
refer to \citet{Rasm06, San03, Fang}. Other more theoretical references dedicated to asymptotic properties are \citet{Stein99} and \citet{Bach14}.

\subsection{Calibration using a GPE}
In Equation (\ref{MC}), the simulations $y_{\ms{\theta}}(\mb{X}^f)$ are replaced with a GPE constructed from a design of experiments $\mb{D}_M$. Let,
\begin{itemize}
\item
$m_{\ms{\beta}}(\mb{D}_M)=\big(h(\mb{x}^{D}_1,\ms{\tau}^{D}_1)^{T}\ms{\beta},\cdots,h(\mb{x}^{D}_M,\ms{\tau}^{D}_M)^{T}\ms{\beta}\big)^{T}$ be the mean vector of the Gaussian process evaluated in each location of $\mb{D}_M$,
\item $m_{\ms{\beta}}(\mb{D}_{\ms{\theta}})$ 
and $\Sigma_{\ms{\Psi}}(\mb{D}_{\ms{\theta}})$ be the 
mean vector and the correlation matrix of the Gaussian process, each  
evaluated in $\mb{D}_{\ms{\theta}}:=\big[(\mb{x}^f_1,\ms{\theta}),\cdots,(\mb{x}^f_n,\ms{\theta})\big)]^{T}\in M_{n,d+p}(\rr),$ 
\item $\Sigma_{\ms{\Psi}}\big(\mb{D}_M,\mb{D}_{\ms{\theta}}\big)$ be the correlation matrix between $\mb{D}_M$ and 
$\mb{D}_{\ms{\theta}}$. 
\end{itemize}
Then, we now consider the available data $\mb{d}:=(y(\mb{D}_M),\mb{z}^f)$. The joint likelihood
of $\ms{\theta}$ and $(\ms{\beta}$, $\sigma^{2}$, $\ms{\Psi})$ is given by

\begin{multline}
\mathcal{L}^{F}(\ms{\theta},\sigma^2,\ms{\beta},\ms{\Psi};\mb{d})\propto
\frac{|C_{\ms{\Psi}}|^{-1/2}}{\sigma^{M+n}}\exp\bigg({-\frac{1}{2\sigma^2}\Big[\big(\mb{d}-(m_{\ms{\beta}}(\mb{D}_M),m_{\ms{\beta}}(\mb{D}_{\ms{\theta}}))}
\big)^{\textrm{T}} \\
C_{\ms{\Psi}}^{-1}\big(\mb{d}-(m_{\ms{\beta}}(\mb{D}_M),m_{\ms{\beta}}(\mb{D}_{\ms{\theta}})
)\big)\Big]\bigg) \,,
\label{totalLike}
\end{multline}
where
$$C_{\ms{\Psi}}=
\begin{pmatrix}
\Sigma_{\ms{\Psi}}(\mb{D}_M) & \Sigma_{\ms{\Psi}}\big(\mb{D}_M,\mb{D}_{\ms{\theta}}\big) \\
\Sigma_{\ms{\Psi}}\big(\mb{D}_M,\mb{D}_{\ms{\theta}}\big)^T & \Sigma_{\ms{\Psi}}(\mb{D}_{\ms{\theta}})+\frac{\lambda^2}{\sigma^{2}}\mb{I}_{n}
\end{pmatrix}\,.
$$

\medskip
In the previous paragraph about the GPE, the parameters $\ms{\beta}$, $\sigma^{2}$ and $\ms{\Psi}$ have been assumed to be known. 
If they are not, their estimation should be conducted jointly with $\ms{\theta}$ based on the full likelihood (\ref{totalLike}) \citep{Higd04}. 
However, inspired by the pioneering work of \citet{Koh2001}, a two-step procedure can be conducted instead. 
This technique, known as modularization in \citet{Liu09}, is still used in a recent work dealing with calibration \citep{Gramacy14}.
It first consists in estimating the parameters $\ms{\beta}$, $\sigma^{2}$ and $\ms{\Psi}$  on the basis of only the simulations $y(\mb{D}_M)$
by maximizing the marginal density of $y(\mb{D}_M)$, denoted by $\mathcal{L}^{M}$:

\begin{multline}
\label{LM}
\mathcal{L}^M(\sigma^2,\ms{\beta},\ms{\Psi};y(\mb{D}_M))\propto\frac{|\Sigma_{\ms{\Psi}}(\mb{D}_M)|}{\sigma^M}^{-1/2}
\exp{\bigg\{-\frac{1}{2\sigma^2}\Big[(y(\mb{D}_M) }-
m_{\ms{\beta}}(\mb{D}_M))^{\textrm{T}} \\
\Sigma_{\ms{\Psi}}(\mb{D}_M)^{-1}(y(\mb{D}_M)-m_{\ms{\beta}}(\mb{D}_M))^{\textrm{T}}\Big]\bigg\}\,.
\end{multline}
Then, the $\mathcal{L}^{M}$ maximum likelihood estimates (MLEs) $(\hat{\sigma}^2,\hat{\ms{\beta}},\hat{\ms{\Psi}})$ of $(\sigma^2,\ms{\beta},\ms{\Psi})$ 
are plugged into the likelihood of $\mb{z}^f$ conditional to the simulations $y(\mb{D}_M)$, denoted below by $\mathcal{L}^C$:
\begin{multline}
\label{eq8}
\mathcal{L}^C(\ms{\theta};\mb{z}^f|y(\mb{D}_M))
\propto|V_{\ms{\hat{\Psi}},\hat{\sigma}^2}^{M}(\ms{\theta})+\lambda^2\mb{I}_n|^{-1/2}\exp{\bigg\{-\frac{1}{2}\Big[(\mb{z}^f-
\mu^{M}_{\hat{\ms{\beta}},\hat{\ms{\Psi}}}(\mb{D}_{\ms{\theta}}))^{\textrm{T}})}\\
(V_{\ms{\hat{\Psi}},\hat{\sigma}^2}^{M}(\ms{\theta})+\lambda^2\mb{I}_n)^{-1}(\mb{z}^f-\mu^{M}_{\hat{\ms{\beta}},\hat{\ms{\Psi}}}(\mb{D}_{\ms{\theta}}))
\Big]\bigg\}
\end{multline}
where
$$\mu_{\ms{\hat{\beta}},\hat{\ms{\Psi}}}^{M}(\mb{D}_{\ms{\theta}}):=(\mu^{M}_{\hat{\ms{\beta}},\hat{\ms{\Psi}}}(\mb{x}^{f}_{1},\ms{\theta}),\cdots,\mu^{M}_{\hat{\ms{\beta}},\hat{\ms{\Psi}}}(\mb{x}^{f}_{n},\ms{\theta}))^{T},$$
and 
$$V_{\ms{\hat{\Psi}},\hat{\sigma}^2}^{M}(\ms{\theta})(i,j)=\,\,\textrm{Cov}\big(Y_M(\mb{x}^f_i,\ms{\theta}),Y_M(\mb{x}^f_j,\ms{\theta})\big).$$

\medskip
\NNew{\citet{Liu09} provide evidence for estimating $\ms{\theta}$ and the set of parameters $\sigma^2$, $\ms{\beta}$ and $\ms{\Psi}$ separately when calibrating a code.
In particular, they demonstrated a poorer mixing in the MCMC routine based on the full likelihood (\ref{totalLike}) than based on (\ref{eq8}).} 
\citet{Koh2001} argue it is not a great loss to estimate $\sigma^2$, $\ms{\beta}$ and $\ms{\Psi}$ only from $y(\mb{D}_M)$ because the number of field measurements $n$
is usually much smaller than $M$.
\citet{Bay07} mentioned cases where the parameters of the GPE may also tune the model in addition to the parameter $\ms{\theta}$ and then \NNew{confounding
between the GPE parameters and the parameter $\ms{\theta}$ could occur.}

From now on, the conditional likelihood (\ref{eq8}) is referred to as the surrogate likelihood. Let $\pi^{C}$ denote the surrogate posterior distribution (SPD) induced by (\ref{eq8}). Then,
\begin{equation}
\label{Bayescalib}
\pi^{C}(\ms{\theta}|\mb{z}^f,y(\mb{D}_M))\propto \mathcal{L}^C(\ms{\theta};\mb{z}^f|y(\mb{D}_M))\pi(\ms{\theta}).
\end{equation}
The SPD (\ref{Bayescalib}) and the APD (\ref{eq4}) are different in that $y_{\ms{\theta}}(\mb{X}^f)$ is replaced by the mean vector of the GPE  $\mu_{\ms{\beta}}^{M}(\mb{D}_{\ms{\theta}})$
and the conditional covariance matrix $V_{\ms{\hat{\Psi}},\hat{\sigma}^2}^M(\ms{\theta})$ is added up to $\lambda^2 \mb{I}_n$. 
Unlike the APD, the SPD is cheap to evaluate up to the normalizing constant, enabling to perform an MCMC algorithm to estimate $\ms{\theta}$.  

\begin{Rem}
The first stage of the modular approach neglects the uncertainty of the parameters $\sigma^2$, $\ms{\beta}$ and $\ms{\Psi}$ by fixing them to their MLE. However, a possible manner to take into account their uncertainty would consist in adopting a Bayesian inference of $\sigma^2$, $\ms{\beta}$ and $\ms{\Psi}$ 
in the same way as $\ms{\theta}$. For instance, if a Jeffreys prior distribution is specified on $(\ms{\beta},\sigma^2)$, then the marginal distribution of the GPE
will follow a Student distribution once $(\ms{\beta},\sigma^2)$ have been integrated out \citep{San03}. Yet unfortunately, the conditional likelihood (\ref{eq8})
has no further closed-form expression, causing an additional issue which is beyond the scope of this paper.
\end{Rem}

\paragraph*{In presence of a code discrepancy $b(\mb{x})$} The calibration setting (\ref{eq4}) has to be modified to prevent overfitting of $\ms{\theta}$, as showed in \citet{Bay07}. In \citet{Higd04} and \citet{Bay07}, 
$b(\mb{x})$ is modeled by a zero mean Gaussian process 
\begin{equation}
\label{GPprior_b}
b(.)\thicksim\mathcal{GP}(0,\sigma_b^2 C_{\ms{\Psi}_b}(.,.))
\end{equation}
where $(.)$ corresponds here to an input $\mb{x}$. In this case, 
{the likelihood arising from the marginal density of $z^f\mid y(D^M)$ once $b(X^f)$ has been integrated out is:
\begin{equation}
\label{likebias}
\mathcal{L}(\ms{\theta};\mb{z}^f)\propto
\frac{1}{|V^f_b+\lambda^2\mb{I}_n|^{1/2}}\exp\Big[{-\frac{1}{2\lambda^2}SS_b(\ms{\theta})\Big]},
\end{equation}
where}
{the sum of squares given in Equation (\ref{MC}) has been replaced with:
}
\begin{equation}
 \label{MCP}
 SS_b(\ms{\theta})=(\mb{z}^f-y_{\ms{\theta}}(\mb{X}^f))^{T}(V^f_b+\lambda^2\mb{I}_n)^{-1}(\mb{z}^f-y_{\ms{\theta}}(\mb{X}^f))
\end{equation}
where $V^f_b(i,j)=\sigma_b^2 C_{\ms{\Psi}_b}(\mb{x}^f_i,\mb{x}^f_j)$. 
Sometimes, the physical context helps us to fix both $\sigma_b^2$ and $\ms{\Psi}_b$ to plausible values, as done in \citet{Craig01}. In such cases, $V^f_b$ becomes known 
and the algorithms that we propose in Section \ref{sec:three_calib} will still be practicable based on (\ref{MCP}) instead of (\ref{MC}). 

\medskip
\paragraph*{Main goal of the paper} Our work focuses on reducing the distance between the SPD (\ref{Bayescalib}) and the APD (\ref{eq4}). The Kullback-Leibler (KL) divergence shows interesting theoretical properties to measure how far a probability distribution is from a reference one \citep{Cover06}. It is written as
\begin{equation}
\label{KLdiv}
\textrm{KL}\big(\pi(\ms{\theta}|\mb{z}^f)||\pi^{C}(\ms{\theta}|\mb{z}^f,y(\mb{D}_M))\big)
=\int_{\mathcal{T}}\pi(\ms{\theta}|\mb{z}^f)\Big(\log{(\pi(\ms{\theta}|\mb{z}^f))}-\log{(\pi^{C}(\ms{\theta}|\mb{z}^f,y(\mb{D}_M))\Big)}\dd\ms{\theta}.
\end{equation}

By using results of approximation theory, we can prove the proposition below.
\begin{Prop}
\label{PropKL}
Under the following assumptions:
\begin{enumerate}
\item[A1] $\pi(\ms{\theta})$ has a bounded support $\mathcal{T}$, 
\item[A2] the code output $y_{\ms{\tau}}(\mb{x})$ is uniformly bounded on $\mathcal{T}\times\mathcal{X}$,
\item[A3] the correlation function (kernel) is a classical radial basis function \citep{Scha95b} {i.e. there exists a function $\phi$ such that $C_{\ms{\Psi}}((\mb{x}',\ms{\tau}'),(\mb{x},\ms{\tau}))=\phi(-\Vert (\mb{x}',\ms{\tau}')-(\mb{x},\ms{\tau}) \Vert)$ where
$\Vert\cdot \Vert$ can be chosen as the Euclidean norm},
\item[A4] the function $y$ lies in the Reproducing Kernel Hilbert Space
associated with the kernel defining the correlation function,
\item[A5] the covering distances associated with the sequence of designs $(\mb{D}_M)_M$:
\begin{equation}
\nonumber
h_{\mb{D}_M}=\max_{(\mb{x}',\ms{\tau}')\in\mathcal{X}\times\mathcal{T}}\min_{(\mb{x}^D_i,\ms{\tau}^D_i)\in \mb{D}_M}\Vert 
(\mb{x}',\ms{\tau}')-(\mb{x}_i,\ms{\tau}_i)\Vert\underset{M\to\infty}\longrightarrow 0\,.
\end{equation}
\end{enumerate}
then, we have:
 \begin{equation}
\label{robust}
\lim_{M\to\infty}\textrm{KL}\big(\pi(\ms{\theta}|\mb{z}^f)||\pi^{C}(\ms{\theta}|\mb{z}^f,y(\mb{D}_M))\big)=0\,.
\end{equation}
\end{Prop}
\textit{Proof.}
(\ref{robust}) results from the uniform convergence of $|\log{(\pi(\ms{\theta}|\mb{z}^f))}-\log{(\pi^{C}(\ms{\theta}|\mb{z}^f,y(\mb{D}_M))}|$ to $0$ 
over $\mathcal{T}$ when $M$ tends to $\infty$ (see Appendix). 
Then, we can exchange limit and integral in (\ref{KLdiv}), which completes the proof.

\medskip
{Assumptions A1-A5 are quite standard. {As} the input domain
of $y$ is usually bounded, then a bounded support prior distribution can be chosen for $\pi(\ms{\theta})$ (A1) and $y$ is assumed to be uniformly bounded
on a bounded domain (A2). Standard choices of kernels such as the 
Gaussian or Matérn kernels are radial basis functions (A3)
and Assumption A4 links the regularity of
the function $y$ with the choice of kernel.
}
{This proposition establishes that the calibration based on the SPD (\ref{Bayescalib}) will be as close as wanted to
the APD (\ref{eq4}) provided that 
$D_M$ grows in such a way that the distances $h_{\mb{D}_M}$ tend to zero (A5).}
However, when $M$ is small with respect to the dimension of the input space $\mathcal{X}\times\mathcal{T}$, (\ref{Bayescalib}) can be significantly different from (\ref{eq4}) leading to a large KL divergence (\ref{KLdiv}). Although a SPD constructed from an accurate GPE is likely to yield a small value of the KL divergence (\ref{KLdiv}), our own experience has shown that such behavior is not systematic.

\paragraph*{\NNew{A heuristic for constructing a design of experiments $\mb{D}_M$ adapted to the GPE-based calibration}}

In practical use, $\mb{D}_M$ is often constructed as a space-filling design (SFD) on the input space $\mathcal{X}\times\mathcal{T}$ ({see \citet{Pronz12} for a deep review of SFD}),
and hence it does not take into account that {$\mb{x}$ and $\ms{\tau}$ play a different role in the computer code}.
Actually, our goal is not to accurately predict the computer code over $\mathcal{X}\times\mathcal{T}$ but to minimize the KL divergence (\ref{KLdiv}). It can be developed as

\begin{multline}
\label{KL_expression_dev}
\textrm{KL}(\pi(\ms{\theta}|\mb{z}^f)||\pi^{C}(\ms{\theta}|\mb{z}^f,y(\mb{D}_M))=
\underbrace{K-K_M}_{(A)}+
\int_{\mathcal{T}}\pi(\ms{\theta}|\mb{z}^f)\underbrace{(C-C_M(\ms{\theta}))}_{(B)}\dd\ms{\theta}
\\+\frac{1}{2}\int_{\mathcal{T}}\pi(\ms{\theta}|\mb{z}^f)
\underbrace{\Big((\mb{z}^f-
\mu^{M}_{\ms{\hat{\beta}},\ms{\hat{\Psi}}}(\mb{D}_{\ms{\theta}}))^{\textrm{T}}
(V^{M}_{\ms{\hat{\Psi}},\hat{\sigma}^2}(\ms{\theta})+\lambda^2\mb{I}_n)^{-1}(\mb{z}^f-\mu^{M}_{\ms{\hat{\beta}},\ms{\hat{\Psi}}}(\mb{D}_{\ms{\theta}}))-SS(\ms{\theta})/\lambda^2\Big)}_{(C)}\dd\ms{\theta}
\end{multline}
where $K$ and $K_M$ correspond to the normalizing constants:
\begin{equation}
K=-\log\left(\int_{\mathcal{T}}\mathcal{L}(\ms{\theta};\mb{z}^f)\pi(\ms{\theta})\right),\quad K_M=-\log\left(\int_{\mathcal{T}}
\mathcal{L}^C(\ms{\theta};\mb{z}^f|y(\mb{D}_M))\pi(\ms{\theta})\right)
\end{equation}
and
\begin{equation}
C=-\frac{n}{2}\log{\lambda^2},\quad C_M(\ms{\theta})=-\frac{1}{2}\log{|V^{M}_{\ms{\hat{\Psi}},\hat{\sigma}^2}(\ms{\theta})+\lambda^2\mb{I}_n|}.
\end{equation}

\NNew{In Equation (\ref{KL_expression_dev}), we decompose the KL
divergence into three parts between terms related to the APD and ones related to the SPD. 
The difference (A), since it is a log ratio of integrated likelihoods over the prior distribution, does not offer much convenient tuning of the design. Our intuition is that focusing on the differences (B) and (C) to construct $\mb{D}_M$ is sufficient.
Since both differences are integrated over $\mathcal{T}$ and
weighted by the APD: $\pi(\ms{\theta}|\mb{z}^f)$, the smaller they are at locations $\ms{\theta}\in\mathcal{T}$ where the APD is high, the more their integral is reduced. 
Our heuristic then suggests to seek for a design $\mb{D}_M$ which contains locations in the form $(\mb{x}^f_i,\ms{\tau})_{1\le i\le n}$ 
where the $\ms{\tau}$ coordinate is chosen in regions where the APD is high.
Indeed, for a given $\ms{\tau}$}
\NNNew{such that $\{(\mb{x}^f_1,\ms{\tau} ),\ldots,(\mb{x}^f_n,\ms{\tau} )\}\subset \mb{D}_M$, 
since $V^{M}_{\ms{\hat{\Psi}},\hat{\sigma}^2}(\ms{\tau})=0$ and $\mu^{M}_{\hat{\ms{\beta}},\hat{\ms{\Psi}}}(\mb{x}^{f}_{i},\ms{\tau})=y_{\ms{\tau}}(\mb{x}^f_i)$
for $i=1,\ldots,n$
from Equation \eqref{eq:interpolator}, the differences (B) and (C) cancel out and 
they remain close to $0$ for any $\ms{\tau}'$ in the neighborhood of $\ms{\tau}$ by the regularity properties of the GPE.
}

\NNew{
Such a design $\mb{D}_M$ can actually be obtained as a natural by-product of a sequential and global maximization procedure for searching $\max_{\ms{\theta}} \pi(\ms{\theta}|\mb{z}^f)$. This procedure allocates the budget of simulation between
locations where the APD is high with respect to the $\ms{\tau}$ coordinate
and ones where exploration is needed. By doing so, the code is likely to have been run over values of $\ms{\tau}$ which lie mainly in all the regions where the APD is high.
By using the log scale and neglecting terms which do not depend on $\ms{\theta}$, the maximization problem is equivalent to solving 
\begin{equation}
\label{cost_with_prior}
 \max_{\ms{\theta}} -SS(\ms{\theta})/2\lambda^2+\log(\pi(\ms{\theta}))\,.
\end{equation}
When little knowledge is available on the value of $\ms{\theta}$, either a uniform prior
(if both a lower and an upper bound are provided) or a locally uniform prior is usually specified for $\ms{\theta}$ \citep{BoxTiao73}.}
\NNNew{
In such cases, when there is substantial information in the data, 
the regions of high probability for $\pi(\ms{\theta}|\mb{z}^f)$ are where $SS(\ms{\theta})$ is small.
In the next section, we present our algorithms for constructing $\mb{D}_M$ in these cases. They are therefore based on
the sequential minimization of $SS(\ms{\theta})$. Hence, the construction of the design $\mb{D}_M$ will be independent on the value of $\lambda^2$.
When the likelihood on $\ms{\theta}$ is flat or if an informative prior is available,
the construction of the design can be based on the optimization problem (\ref{cost_with_prior}) which takes into account the prior at no additional cost.
In this latter case, the construction of the design will depend on the the value of $\lambda^2$ since it balances the weight given to the sum of squares and the one given to the prior.
}
\NNew{
\begin{Rem}
We aim to construct sequential designs through the minimization of $SS(\ms{\theta})$ in order to perform GPE-based Bayesian calibration.
Our work is different from those of \citet{Joseph09} and \citet{Wong14} which deal with calibration as a pure optimization problem of $SS(\ms{\theta})$.
\end{Rem}
}


%

%



\section{Adaptive designs for calibration}

\label{sec:three_calib}

To identify the global minimum of a costly black box code, denoted by $f$ (to avoid confusion with $y$ in the calibration setting), Expected Improvement (EI) criterion-based strategies can be performed \citep{Jones98}. They consist in identifying sequentially 
the input locations where the code $f$ should be run to be close to the global minimum, which is relevant when only a small number of simulations is allocated.
Assuming $k$ simulations $f(\mb{D}_{k})$ have already been run, the EI criterion assesses the expected
improvement of a new run (numbered $k+1$) in terms of getting close to the unknown global minimum of $f$. Let $\mb{v}_{k+1}$ be the input 
where the EI value is at its highest:
\begin{align}
\label{eq9}
\nonumber
\mb{v}_{k+1}=&\,\,\,\argmax{\mb{v}}{EI_{k}(\mb{v})},\\
=&\,\,\,\argmax{\mb{v}}{\ee[(m_k-F_{k}(\mb{v}))\mb{1}_{F_{k}(\mb{v})<m_k}]}\,,
\end{align}
where 
\begin{itemize}
\item $F_k(\mb{v})$ is the current GPE which is built from $\mb{D}_k$,
\item $m_k=\min{\{f(\mb{v}_1),\cdots,f(\mb{v}_{k-1}),f(\mb{v}_k)\}}$ is the current value for the minimum.
\end{itemize}
If a deterministic emulator were used instead of $F_{k}(\mb{v})$, for instance the mean $\mu(\mb{v})$ of $F_{k}(\mb{v})$, the EI criterion would just be the difference $m_k-\mu(\mb{v})$ if $\mu(\mb{v})<m_k$ and $0$ if $\mu(\mb{v})>m_k$. Given that $F_{k}(\mb{v})$ is stochastic, Equation (\ref{eq9}) is written as the expectation of this truncated difference with respect to the distribution of $F_{k}$.
The algorithm that consists of running the code at the input $\mb{v}_{k+1}$ then updating the emulator and repeating this 
is called Efficient Global Optimization (EGO) \citep{Jones98}. The convergence of the EGO algorithm to the global minimum of $f$ has been proven with respect to some assumptions about both the smoothness of the code and the correlation function of the GPE \citep{Vasq10}. 
In its current use, the algorithm is stopped either when the number of allocated simulations is exceeded or when the improvement of $m_k$ becomes negligible. According to this last criterion, EGO requires less simulations than other optimization methods with comparable levels of performance \citep{Ginsbourger}. 

\medskip
\paragraph*{EI designed for calibration} Our contribution now consists in resorting to
the EI criterion for the sum of squares of the residuals function $SS(\ms{\theta})$ {(defined in Equation (\ref{MC}))}:
\begin{equation}
\label{EI_calib}
EI_{k}(\ms{\theta})=\ee \left[\big(m_k-SS_{k}(\ms{\theta})\big)\mb{1}_{SS_{k}(\ms{\theta})\leq m_k}\right]\,\,\,\in\,\,\,[0,m_k],
\end{equation}
where 
\begin{itemize}
\item $m_k:=\min{\{SS(\ms{\theta}_1),\cdots,SS(\ms{\theta}_{k-1}),SS(\ms{\theta}_k)\}}$ and $SS(\cdot)$ denotes the sum of squares computed from actual runs of the computer code $y$,
\item \NNew{$SS_{k}(\ms{\theta})$ denotes the sum of squares of the residuals where $y_{\ms{\theta}}(\mb{X}^{f})$ is replaced with the random vector
$Y_{k}(\mb{D}_{\ms{\theta}})=\big(Y_k(\mb{x}^f_1,\ms{\theta}),\cdots,Y_k(\mb{x}^f_n,\ms{\theta})\big)$, the distribution of which is given by the GPE conditional to $y(\mb{D}_k)$:
$$ SS_{k}(\ms{\theta})=||\mb{z}^f-Y_{k}(\mb{D}_{\ms{\theta}})||^{2}\,.$$}
\end{itemize}
Note that the subscript $k$ now refers to the current iteration of the algorithm.
$SS_{k}(.)$ is thus a random process and its distribution inherits from the current GPE. At step $k$, \NNew{$n$} new simulations need to be run to update $m_k$. Hence, the design $\mb{D}_k$ contains all the simulations $y_{\ms{\theta}_j}(\mb{x}^f_i)$ for all $1\le i\le n$ and $1\le j\le k$ (should not be mistaken for the notation in Section \ref{sec:two_calib} where $M$ has referred to the number of simulations). Let $\ms{\theta^{\star}}$ be the maximum of (\ref{EI_calib}):
\begin{equation*}
\label{maxi}
\ms{\theta^{\star}}=\argmax{\ms{\theta}}{EI_{k}(\ms{\theta})}.
\end{equation*}

%
%

\paragraph*{EGO algorithms} Algorithm $1$ corresponds to an exact EGO algorithm based on Equation (\ref{EI_calib}). It aims to identify the input locations $(\mb{x}^f_i,\ms{\theta}^{\star})\in\mb{X}^f\times\mathcal{T}$ which will be added up sequentially to an initial design $\mb{D}_0$ for $M$ iterations. Algorithm $2$ is a one at time algorithm and should be understood as an approximation of Algorithm $1$.

\bigskip
\vspace{3pt}\hrule\vspace{6pt}
\noindent Algorithm $1$\par\nobreak
\vspace{3pt}\hrule\vspace{6pt}

\bigskip
\textbf{Initialization}

\begin{itemize}
\item {Choose} an initial numerical design $\mb{D}_0\subset\mathcal{X}\times\mathcal{T}$ of size $M_0$.

\item Run the code over $\mb{D}_0$, {then construct an initial GPE based on $y(\mb{D}_{0})$}.

\item {Compute $\ms{\hat{\ms{\theta}}}_1$ as the posterior mean $\ee[\ms{\theta}|\mb{z}^f,y(\mb{D}_0)]$}.

\item $\mb{D}_{1}=\mb{D}_0\cup\{(\mb{x}^f_i,\ms{\hat{\theta}}_{1})\}_{1\leq i \leq n}$.

\item Update the GPE distribution after running the code over $\{(\mb{x}^f_i,\ms{\hat{\theta}}_{1})\}_{1\leq i \leq n}$.

\item {Compute} $m_1:=SS(\ms{\hat{\theta}}_1)$.
\end{itemize}

\smallskip
\textbf{From $k=1$, repeat the following steps as long as $M_{0}+n\times (k+1)\leq M$.}

\bigskip
\textbf{Step $\mb{1}$} Find an estimate $\ms{\hat{\theta}}_{k+1}$ of $\ms{\theta}_{k+1}^{\star}=\argmax{\ms{\theta}}{EI_{k}(\ms{\theta})}$.

\bigskip
\textbf{Step $\mb{2}$} $\mb{D}_{k+1}=\mb{D}_k\cup\{(\mb{x}^f_i,\ms{\hat{\theta}}_{k+1})\}_{1\leq i \leq n}$.

\bigskip
\textbf{Step $\mb{3}$} Run the code over all new locations $\{(\mb{x}^f_i,\ms{\hat{\theta}}_{k+1})\}_{1\leq i \leq n}$.

\bigskip
\textbf{Step $\mb{4}$} Update the GPE distribution based on $y(\mb{D}_{k+1})$.

\bigskip
\textbf{Step $\mb{5}$} {Compute} $m_{k+1}:=\min{\{m_1,\cdots,m_k,SS(\ms{\hat{\theta}}_{k+1})\}}$. 
\vspace{3pt}\hrule\vspace{6pt}

\medskip
{The way we choose} $\mb{D}_0$ will be discussed in Section \ref{sec:four_calib}.

\medskip
Because the distribution of the GPE is updated at Step $4$, the hyper-parameters $\ms{\beta}$, $\sigma^{2}$ and $\ms{\Psi}$
are re-estimated, as done in the seminal work on EGO algorithm \citep{Jones98}.
Algorithm $1$ could be efficiently performed by running the new simulations at Step $3$ simultaneously on several computer nodes. We {lay out} below the steps of a one-at-a-time algorithm well-suited when the computer system has a single processor. 

\bigskip
\vspace{3pt}\hrule\vspace{6pt}
\noindent Algorithm $2$\par\nobreak
\vspace{3pt}\hrule\vspace{6pt}

\bigskip
Initialization is similar to Algorithm $1$ except that
$\mb{D}_{1}=\mb{D}_0\cup\{(\mb{x^{\star}},\ms{\hat{\theta}}_{1})\}$ where $$\mb{x}^{\star}=
\argmax{\mb{x}^f_i\in \mb{X}^f}{\textrm{Crit}(\mb{x}^f_i,\ms{\hat{\theta}}_{1})}\,\,\,
\textrm{(see Equations (\ref{firstcrit}) and (\ref{second_crit}) below)}.$$

\bigskip
\textbf{For} $k=1,\cdots,M-M_{0}$, \textbf{repeat the same steps as in Algorithm $1$ except that 
{Steps $2$, $3$, and $5$ are replaced respectively with Steps $\tilde{2}$, $\tilde{3}$ and $\tilde{5}$.}
}

{\textbf{Step $\mb{\tilde{2}}$} $\mb{D}_{k+1}=\mb{D}_k\cup\{(\mb{x}^*,\ms{\hat{\theta}}_{k+1})\}$ where \\
$$\mb{x}^{\star}=
\argmax{\mb{x}^f_i\in \mb{X}^f}{\textrm{Crit}(\mb{x}^f_i,\ms{\hat{\theta}}_{k+1})}\,\,\, \textrm{(see Equations (\ref{firstcrit}) and (\ref{second_crit}) below)}.$$
}

\bigskip
\textbf{Step $\mb{\tilde{3}}$} Run the code in $(\mb{x^{\star}},\ms{\hat{\theta}}_{k+1})$. 

\bigskip
\textbf{Step $\mb{\tilde{5}}$} {Compute} $m_{k+1}:=\min{\{\ee[SS_{k}(\ms{\hat{\theta}}_{1})],\cdots,\ee[SS_{k}(\ms{\hat{\theta}}_{k})],\ee[SS_{k}(\ms{\hat{\theta}}_{k+1})]\}}$

\vspace{3pt}\hrule\vspace{6pt}

\bigskip
{Unlike Algorithm $1$, a single simulation $y_{\ms{\hat{\theta}}_{k+1}}(\mb{x}^{\star})$ is run at each iteration in Algorithm $2$. As the current minimum cannot thus be computed anymore, we have replaced it by the minimum of the expectation values $\ee[SS_{k}(\ms{\hat{\theta}}_i)]$ for $1\le i \le k+1$ that are taken with respect to $Y_{k}(.)$.}
In what follows, two criteria are proposed to choose a relevant $\mb{x}^{\star}$ among $\mb{X}^f$ in Step $\mb{\tilde{3}}$. 

The first criterion aims to select the input location $(\mb{x^{\star}},\ms{\hat{\theta}}_{k})\in \mb{X}^f\times\mathcal{T}$ where the variance of $Y_{k}(.)$ is at the highest, 

\begin{equation}
\label{firstcrit}
\textrm{Crit}(\mb{x}^f_i,\ms{\hat{\theta}}_{k+1})=\vv[Y_{k}(\mb{x}^f_i,\ms{\hat{\theta}}_{k+1})]\,.
\end{equation}
As the variance of the Gaussian process decreases on the space $\mb{X}^f\times\mathcal{T}$, the SPD (\ref{Bayescalib})
gets closer to the APD (\ref{eq4}), which justifies (\ref{firstcrit}). Yet, a better way might perhaps consist in aiming for a reduction of the GPE uncertainty at an input location $(\mb{x^{\star}},\ms{\hat{\theta}}_{k+1})$ where the code $y_{\ms{\theta}}(\mb{x}^{\star})$ is highly variable
with respect to $\ms{\theta}$, meaning that $\mb{x}^{\star}$ is influential for calibration. We thus introduce a second criterion which does a trade-off between the calibration goal and (\ref{firstcrit}). A normalized version of it is written as

\begin{equation}
\label{second_crit_exact}
\textrm{Crit}(\mb{x}^f_i,\ms{\hat{\theta}}_{k+1})=\frac{\vv\big(Y_{k}(\mb{x}^f_i,\ms{\hat{\theta}}_{k+1})\big)}
{\max\limits_{i=1,\cdots,n}\vv\big(Y_{k}(\mb{x}^f_i,\ms{\hat{\theta}}_{k+1})\big)}
\times \frac{\vv[y_{\ms{\theta}}(\mb{x}^f_i)]}{\max\limits_{i=1,\cdots,n}\vv[y_{\ms{\theta}}(\mb{x}^f_i)]},
\end{equation}
where $\vv[y_{\ms{\theta}}(\mb{x}^f_i)]$
is taken with respect to $\pi(\ms{\theta})$.
In practice, we need to use an approximation of (\ref{second_crit_exact}) that is based on the mean of $Y_{k}(.)$:

\begin{equation}
\textrm{Crit}(\mb{x}^f_i,\ms{\hat{\theta}}_{k+1})=\frac{\vv\big(Y_{k}(\mb{x}^f_i,\ms{\hat{\theta}}_{k+1})\big)}
{\max\limits_{i=1,\cdots,n}\vv\big(Y_{k}(\mb{x}^f_i,\ms{\hat{\theta}}_{k+1})\big)}
\times \frac{\vv[\mu^{k}_{\ms{\beta},\ms{\Psi}}(\mb{x}^f_i,\ms{\theta})]}{\max\limits_{i=1,\cdots,n}\vv[\mu^{k}_{\ms{\beta},\ms{\Psi}}(\mb{x}^f_i,\ms{\theta})]}\,.
\label{second_crit}
\end{equation}

\NNew{
\begin{Rem}
 Once the new $\ms{\hat{\theta}}_{k+1}$ has been selected, the criteria \eqref{firstcrit} and \eqref{second_crit} for choosing $\mb{x}^{\star}$ in $\mb{X}^f$
 are also derived from the decomposition of the KL divergence \eqref{KL_expression_dev}. The criterion \eqref{firstcrit} is concerned with the variance term
 coming from the GPE which appears in the differences (B) and (C). The criterion \eqref{second_crit} is concerned with both the variance term and the mean term.
 \end{Rem}
}

\begin{Rem}
A typical problem inherent to sequential designs is when two input locations come very close, making the covariance matrix numerically singular
and thus difficult to invert. This issue can arise when both $\ms{\hat{\theta}}_{k}$ is too close to a previous iteration 
$\ms{\hat{\theta}}_{k^{\prime}}$ and $\mb{x}^{\star}$ is almost the same at iterations $k$ and $k^{\prime}$.
The usual way to circumvent it consists in adding a small diagonal matrix to the covariance matrix $V_{\ms{\Psi},\sigma^2}^{k^{\prime}}(\ms{\theta})$ of the GPE, called nugget effect. 
\end{Rem}

\begin{Rem}
\label{RemOAT}
Algorithm $2$ can replace Algorithm $1$ when the simulation budget $M$ is not much larger than $n$. In that case, Algorithm $1$ is unpractical due to the 
impossibility of evaluating $SS(\ms{\theta})$ over a sufficient number of locations $\ms{\theta}_k$.
In \citet{Prat13}, the likelihood ratio is replaced by a GPE, then it is minimized by using sequential runs based on the EI criterion. A similar idea could consist 
in modeling $SS(\ms{\theta})$ as
a GPE.
Both of these methods necessarily need to run $n$ simulations at each iteration to update the distribution 
of the likelihood ratio or that of $SS(\ms{\theta})$ respectively.
Therefore, a one-at-a-time strategy like Algorithm 2 cannot be designed for these methods.
\end{Rem}

\paragraph*{Computation of the EI criterion} By expanding Equation (\ref{EI_calib}), we have : 

\begin{equation}
\label{EI_calib2}
EI_{k}(\ms{\theta})= m_k\Bigg[\pp[SS_{k}(\ms{\theta})<m_k]-\frac{\ee[SS_{k}(\ms{\theta})\mb{1}_{SS_{k}(\ms{\theta})\leq m_k}]}{m_k}\Bigg]>0\,,
\end{equation} 
implying
\begin{equation}
\ee[SS_{k}(\ms{\theta})\mb{1}_{SS_{k}(\ms{\theta)}\leq m_k}]\leq m_k \pp[SS_{k}(\ms{\theta})<m_k].
\end{equation}
Except in the trivial case $n=1$, no closed form expression can be obtained for $EI_{k}(\ms{\theta})$.
{It is calculated within constants by summing up the probability $\pp[SS_{k}(\ms{\theta})<m_k]$ of sampling inside the hypersphere $B(0,\sqrt{m_k})$ from a multivariate 
Gaussian distribution and the expectation of the right truncated $SS_{k}(\ms{\theta})$ with respect to $m_k$}. $\pp[SS_{k}(\ms{\theta})<m_k]$ can be calculated either as an infinite series in central chi-square distribution \citep{S7} or by using an advanced sampling rejection method \citep{Ellis12}. {This second method should be preferably used since the second term cannot be estimated other than using MCMC sampling.} 

The minimization of (\ref{EI_calib2}) can be performed in a greedy fashion where $\ms{\hat{\theta}}_{k+1}$ is taken as the value which maximizes $EI_{k}(\ms{\theta})$ over a grid $G\in\mathcal{T}$. For some candidates of $G$, the computation of $EI_{k}(\ms{\theta})$ could be avoided as explained hereafter.

\bigskip
\vspace{3pt}\hrule\vspace{6pt}
\noindent Computation of $\ms{\hat{\theta}}_{k+1}$\par\nobreak
\vspace{3pt}\hrule\vspace{6pt}
Recall that $Y_{k}(\mb{D}_{\ms{\theta}})$ be the random vector $\big(Y_k(\mb{x}^f_1,\ms{\theta}),\cdots,Y_k(\mb{x}^f_n,\ms{\theta})\big)$.
\begin{enumerate}[1.]
\item Compute $\pp[\mb{z}^f-Y_{k}(\mb{D}_{\ms{\theta}})\in[-m_k,m_k]^{n}]$  which is an upper bound of $\pp[SS_{k}(\ms{\theta})\leq m_k]$ for each $\ms{\theta}$ of $G$,
\item Let $\tilde{\ms{\theta}}=\argmax{\ms{\theta}\in G}\pp[\mb{z}^f-Y_{k}(\mb{D}_{\ms{\theta}})\in[-m_k,m_k]^{n}]$ be a reference value.
\item Compute $EI_{k}(\ms{\tilde{\theta}})$,
\item Build the sub-grid $\tilde{G}=\{\ms{\theta}\in\mathcal{T}\,\,;\,\, EI_{k}(\ms{\tilde{\theta}})\leq\pp[\mb{z}^f-Y_{k}(\mb{D}_{\ms{\theta}})\in[-m_k,m_k]^{n}]\}\subset G$,
\item Compute $EI_{k}(\ms{\theta})$ for the values of the sub-grid $\tilde{G}$,
\item Let $\ms{\hat{\theta}}_{k+1}=\argmax{\theta\in\tilde{G}}{EI_{k}(\ms{\theta})}$.
\end{enumerate}

\bigskip
\noindent
For $\ms{\theta}\in G\setminus \tilde{G}$, 
we have $EI_{k}(\ms{\tilde{\theta}})\geq\pp[\mb{z}^f-Y_{k}\in[-m_k,m_k]^{n}]$ implying
$EI_{k}(\ms{\tilde{\theta}})>EI_{k}(\ms{\theta})$. 
Hence, there is no need to compute $EI_{k}(\ms{\theta})$. 
Unfortunately, this algorithm is only relevant when $n$ is small because in higher dimensions, 
the hypercube $[-m_k,m_k]^{n}$ has a much larger volume than the hypersphere. 

\medskip
Such a greedy optimization works well, especially in small dimensions of $\mathcal{T}$ because $G$ can be constructed fine enough (see Section \ref{sec:four_calib}). In higher dimension, we advise to use several grids that are finer in the region of high probability as iterations occur (see Section \ref{sec:four_calib}). Even if $G$ is coarser, we can expect our algorithms stay efficient in terms of reducing the KL divergence because they do not aim at converging precisely to the global minimum of $SS(\ms{\theta})$, but rather identifying the area of the input space where $SS(\ms{\theta})$ is small.
{Furthermore,
seeking for $\ms{\hat{\theta}}_{k+1}$ on a grid may prevent
from obtaining points in the adaptive design 
which are too close, 
which is often the case when using EI algorithms. 
Thus, 
this will prevent from numerical issues when fitting the GPE.}


 

\section{Simulation study}
\label{sec:four_calib}

\begin{figure}[ht!]
\centering
\caption{\textit{Left: the function $y_{\tau}(x)=(6x-2)^2\times\sin{(\tau x -4)}$ on $[0,1]$ for several 
values of $\tau\in[5,15]$. Red dots are the field measurements $(\mb{X}^f,\mb{z}^f)$ generated by Equation (\ref{data}). 
Right: APD (Case $1$).}}
\includegraphics[scale=0.43]{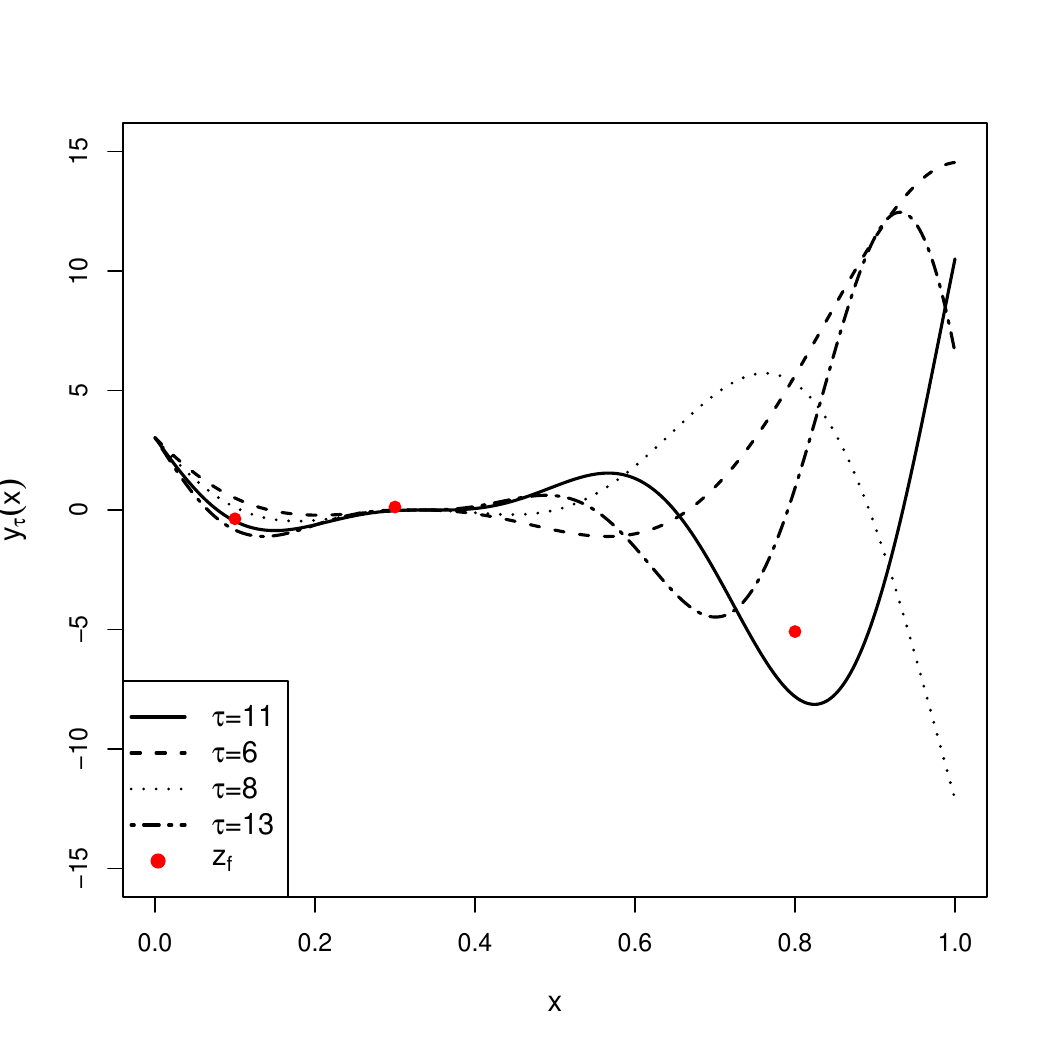}
\includegraphics[scale=0.43]
{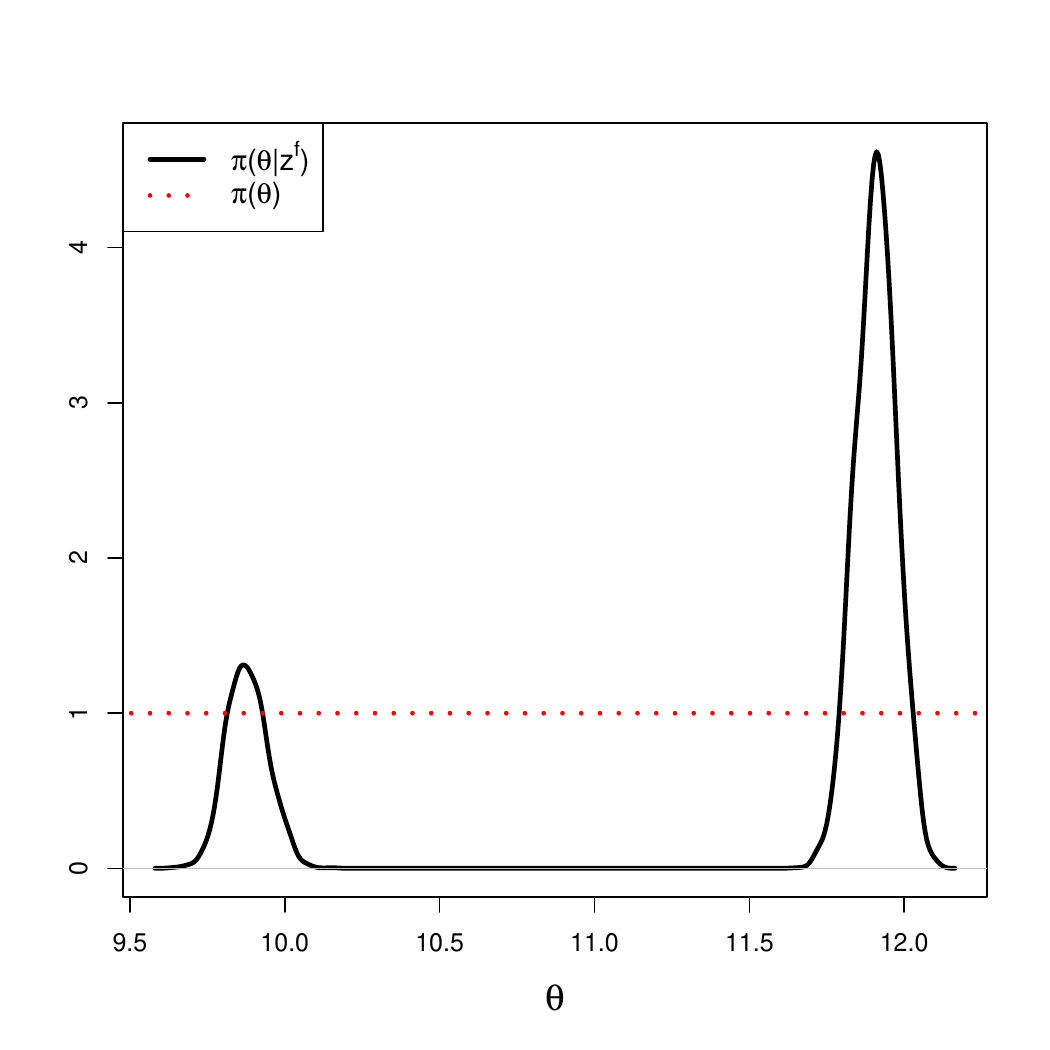}
\label{ftest}
\end{figure}

\paragraph*{A 2D example} Let us assume that the computer code is given by the function:
\begin{equation}
\label{analyticf}
y_{\tau}:x\longrightarrow y_{\tau}(x)=(6x-2)^2\times\sin{(\tau x -4)},
\end{equation}
where $x\in\mathcal{X}=[0,1]$ and $\tau\in\mathcal{T}=[5,15]$. For $1\leq i\leq n$, the field data $\mb{z}^f$ are generated by
\begin{equation}
\label{data}
z^f_i=y_{\theta}(x^f_i)+\epsilon_i,
\end{equation}
where $\theta=12$ and $\epsilon_i\thicksim\mathcal{N}(0,0.3^2)$. Bayesian calibration of the function (\ref{analyticf}) is first performed by sampling 
the APD (\ref{eq4}) (see Figure \ref{ftest}) where the prior distribution
$\pi(\theta)$ is chosen as uniform on $[5,15]$:
\begin{equation}
\pi({\theta})\propto \mb{1}_{[5,15]}(\theta)\,.
\end{equation} 
Then, Bayesian calibration of (\ref{analyticf}) is performed by sampling the SPD (\ref{Bayescalib}). Two cases are addressed: with $n=3$ field measurements, then with $n=9$ field measurements.

\begin{figure}[ht!]
\centering
\caption{\textit{Sampling of SPD (\ref{Bayescalib}) from two different maximin LHD of size $M=30$ (using the R library MCMCpack).}}
\includegraphics[scale=0.43]{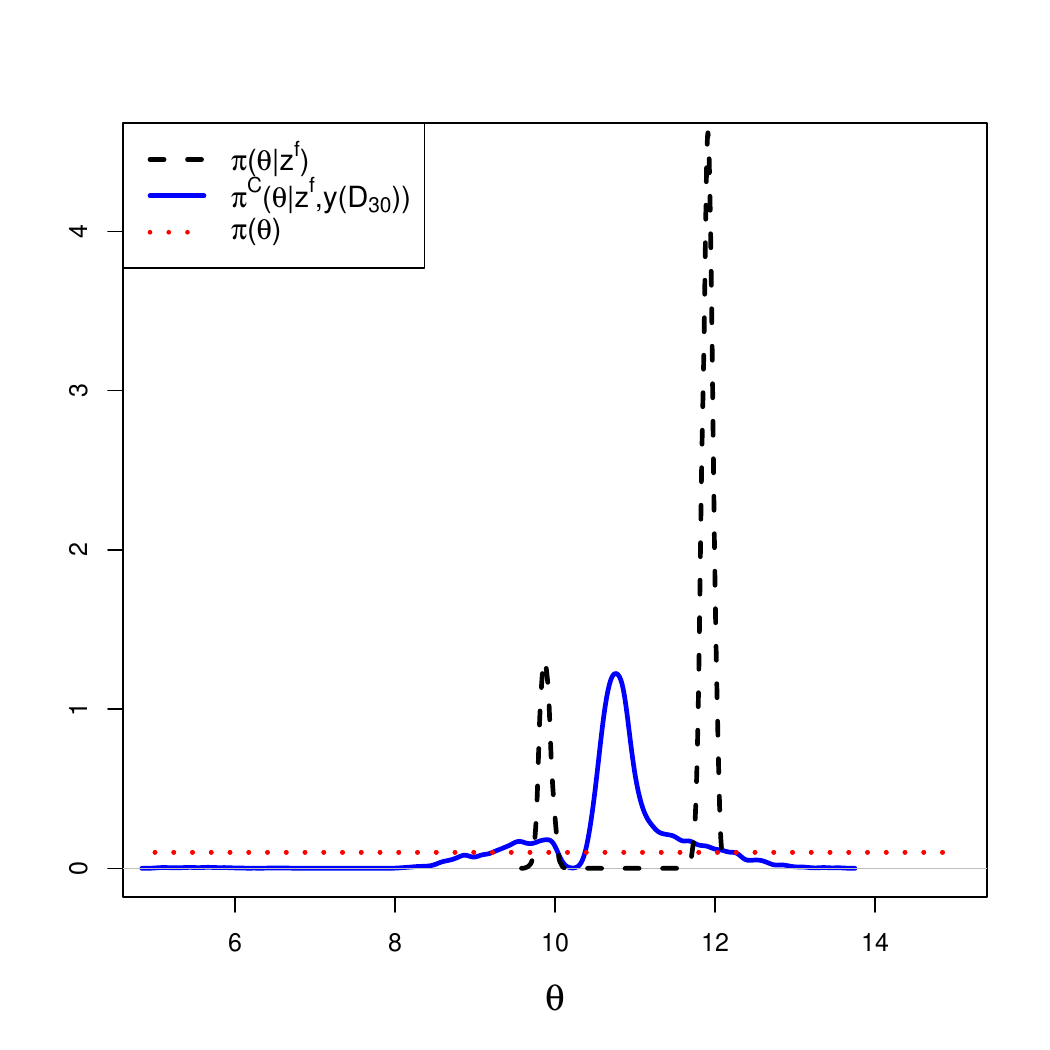}
\includegraphics[scale=0.43]
{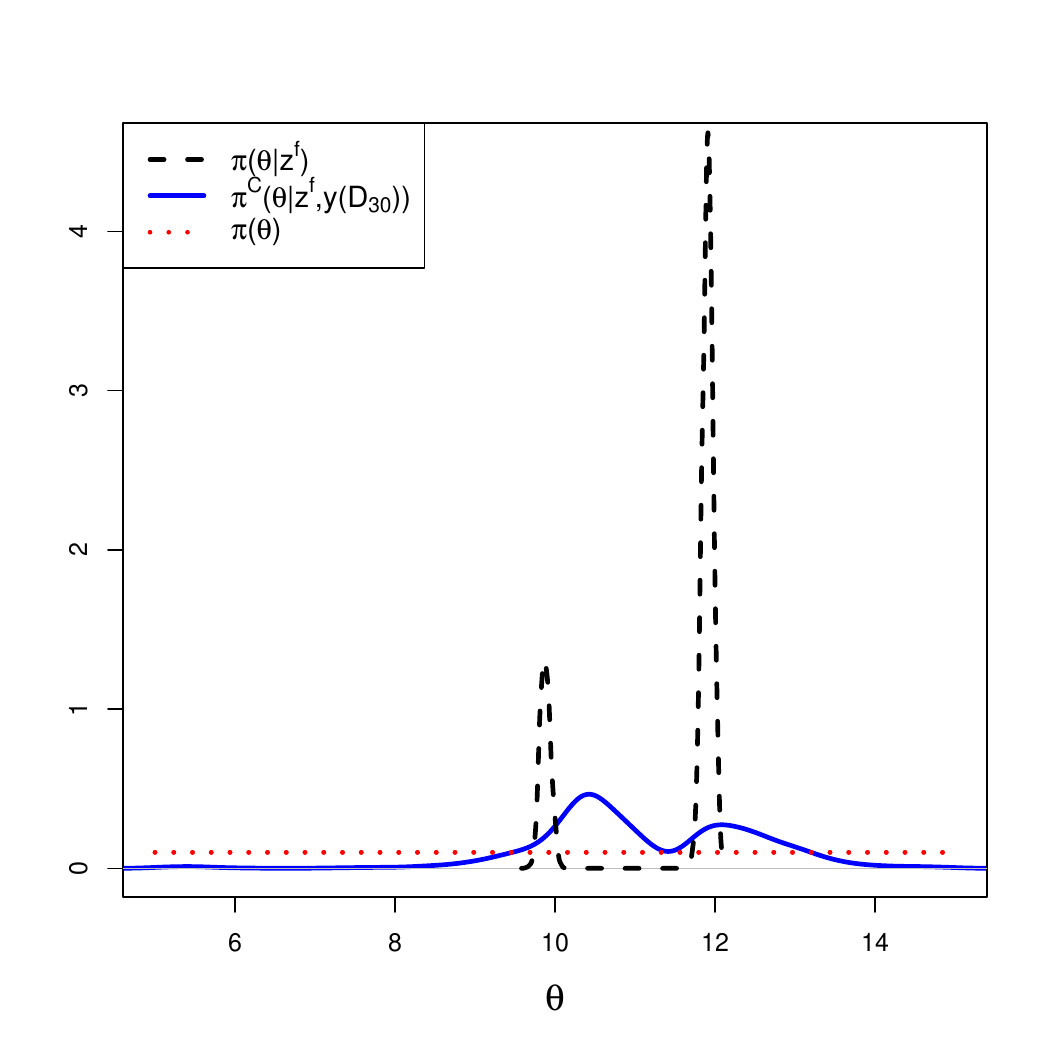}
\label{MaxiNR}
\end{figure}

\paragraph*{Case $\mb{1}$: $\mb{X}^f=(0.1,0.3,0.8)$} 
The SPD 
(\ref{Bayescalib}) is constructed with a GPE having a constant mean  $m_{\ms{\beta}}=m$ and a Matérn $5/2$ correlation function. 
All the parameters $m,\sigma^{2},\Psi$ are estimated using the modular approach described in Section \ref{sec:two_calib}.
Figure~\ref {MaxiNR} shows the results obtained when the GPE is estimated from two different one-shot maximin LHD. On the right-hand-side, the regions of high posterior density roughly match with those of the APD but the two modes 
are reversed in terms of height. On the left-hand-side, the SPD looks strongly different from the actual one. Such GPE-based calibrations need to be much improved. 

In Section \ref{sec:two_calib}, we have pointed out that the value of the KL divergence only depends on the distribution of the GPE in the subspace $\mb{X}^{f}\times\mathcal{T}$. It is thus relevant to restrict locations of the numerical design of experiments $\mb{D}_M$ to this subspace as adaptive strategies do (see Section \ref{sec:three_calib}). These strategies need to start from an initial design $\mb{D}_0$ with a size that is here set to $M_0=12$ locations and then they will be stopped when $M=30$. Two kinds of $\mb{D}_0$ have been tested including maximin LHD on $\mathcal{X}\times\mathcal{T}$ and \textit{restricted}-to-$\mb{X}^{f}$ maximin LHD, so called because they are maximin on $\mb{X}^{f}\times\mathcal{T}$ with the property that their one-dimensional projections in $\mathcal{T}$ are uniform. As part of this example, let us recall how to construct $\mb{D}_{k+1}$ from $\mb{D}_{k}$ in the adaptive strategies:

\begin{itemize}
\item Strategy $A$: $\mb{D}_{k+1}$ is constructed by adding all the locations $\mb{X}^{f}\times\theta_k$ to the current design $\mb{D}_k$ where $\theta_k$ is the value of $\theta$ maximizing the EI criterion (see Algorithm $1$). 
\item Strategy $B$: $\mb{D}_{k+1}$ is constructed by adding a single location $(x^{\star},\theta_k)$ to the current design $\mb{D}_k$ where $(x^{\star},\theta_k)$ has the highest variance (\ref{firstcrit}) (see Algorithm $2$).
\item Strategy $C$: $\mb{D}_{k+1}$ is constructed by adding a single location $(x^{\star},\theta_k)$ to the current design $\mb{D}_k$ where $(x^{\star},\theta_k)$ maximizes the criterion (\ref{second_crit}) (see Algorithm $2$). 
\end{itemize}

 \begin{figure}[H]
 \centering
\caption{\textit{$\mb{X}^{f}=(0.1,0.3,0.8)^{T}$ Upper left: Design $1$. Upper right: Design $4$. Bottom Left: Design $2$. Bottom right: Design $5$. The black dots are the initial design $\mb{D}_0$. The red stars are the locations sequentially added with the EI criterion.}}

\includegraphics[scale=0.37]{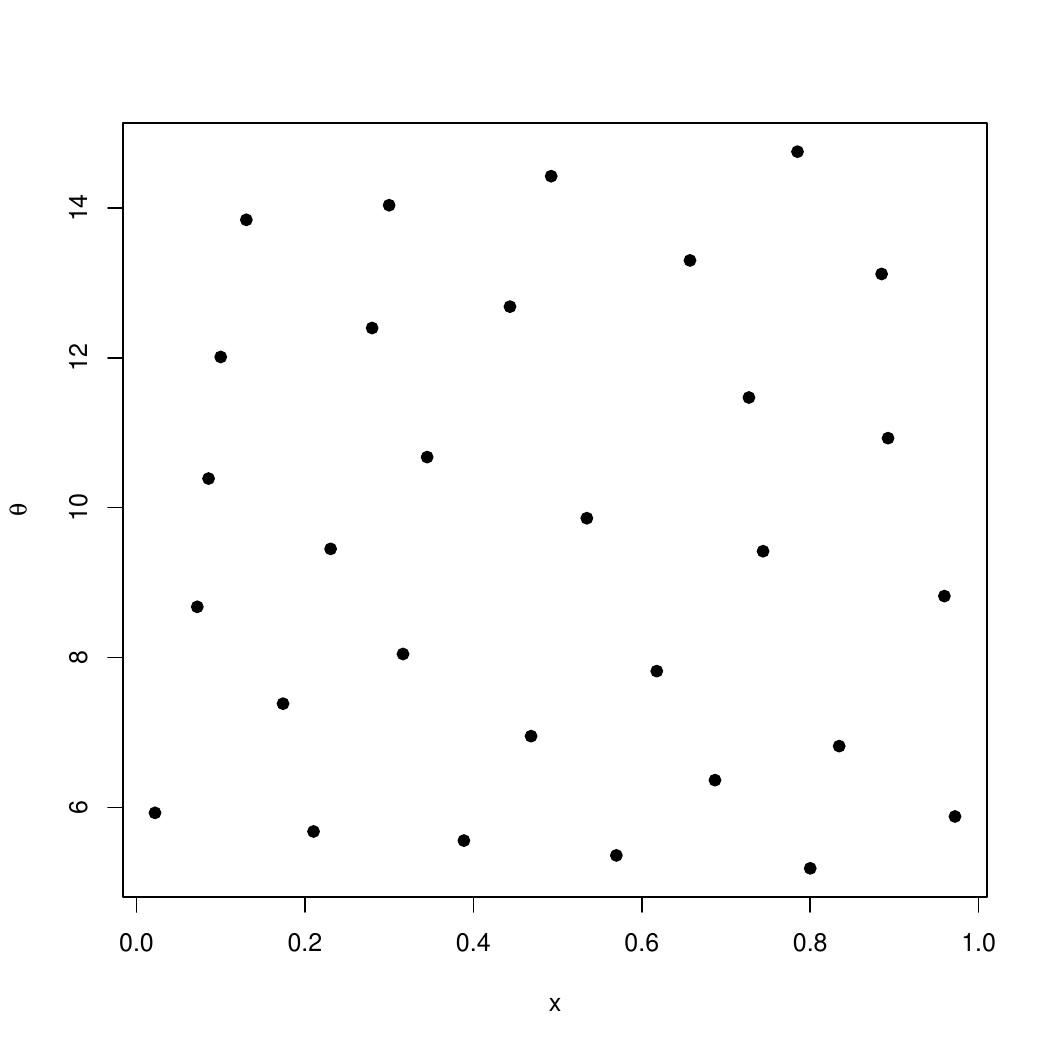}
\includegraphics[scale=0.37]{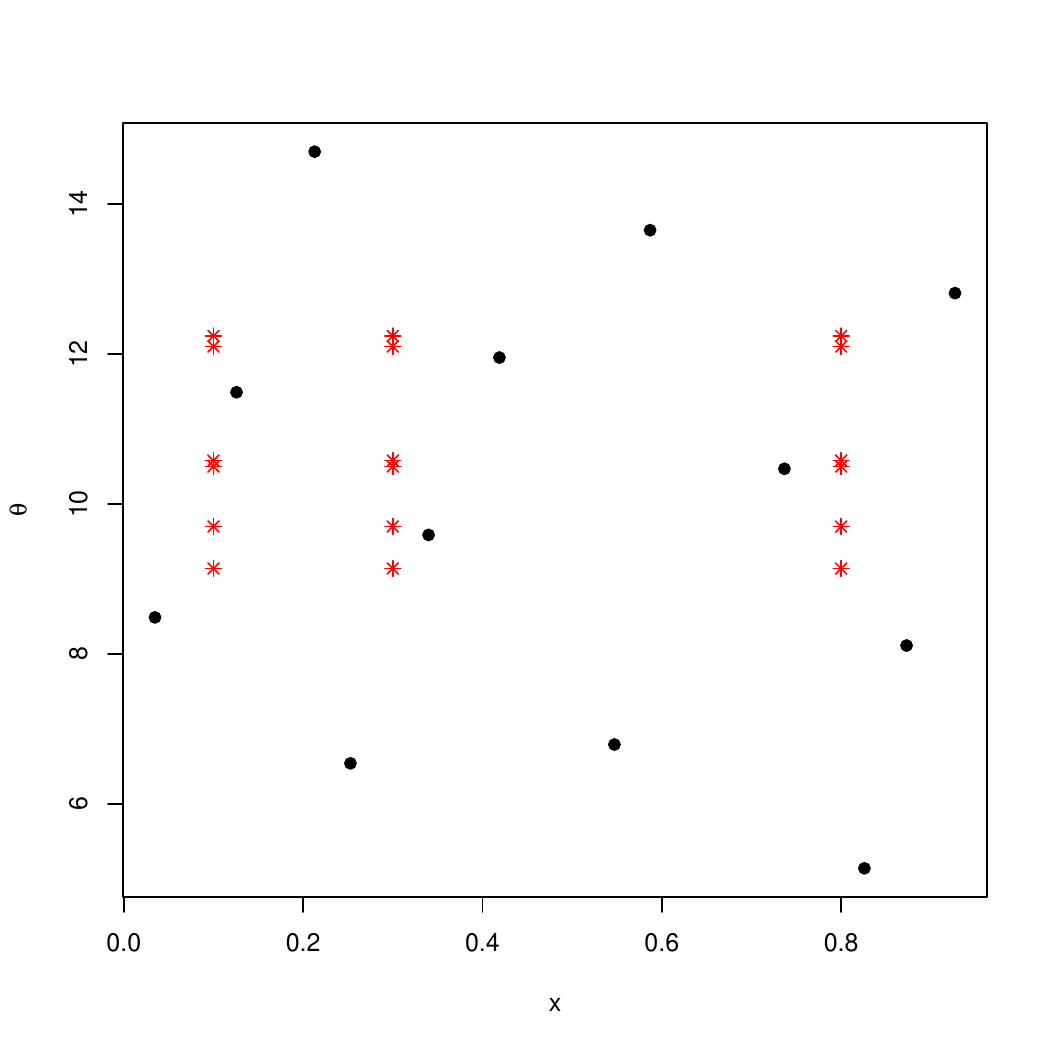}

\includegraphics[scale=0.37]{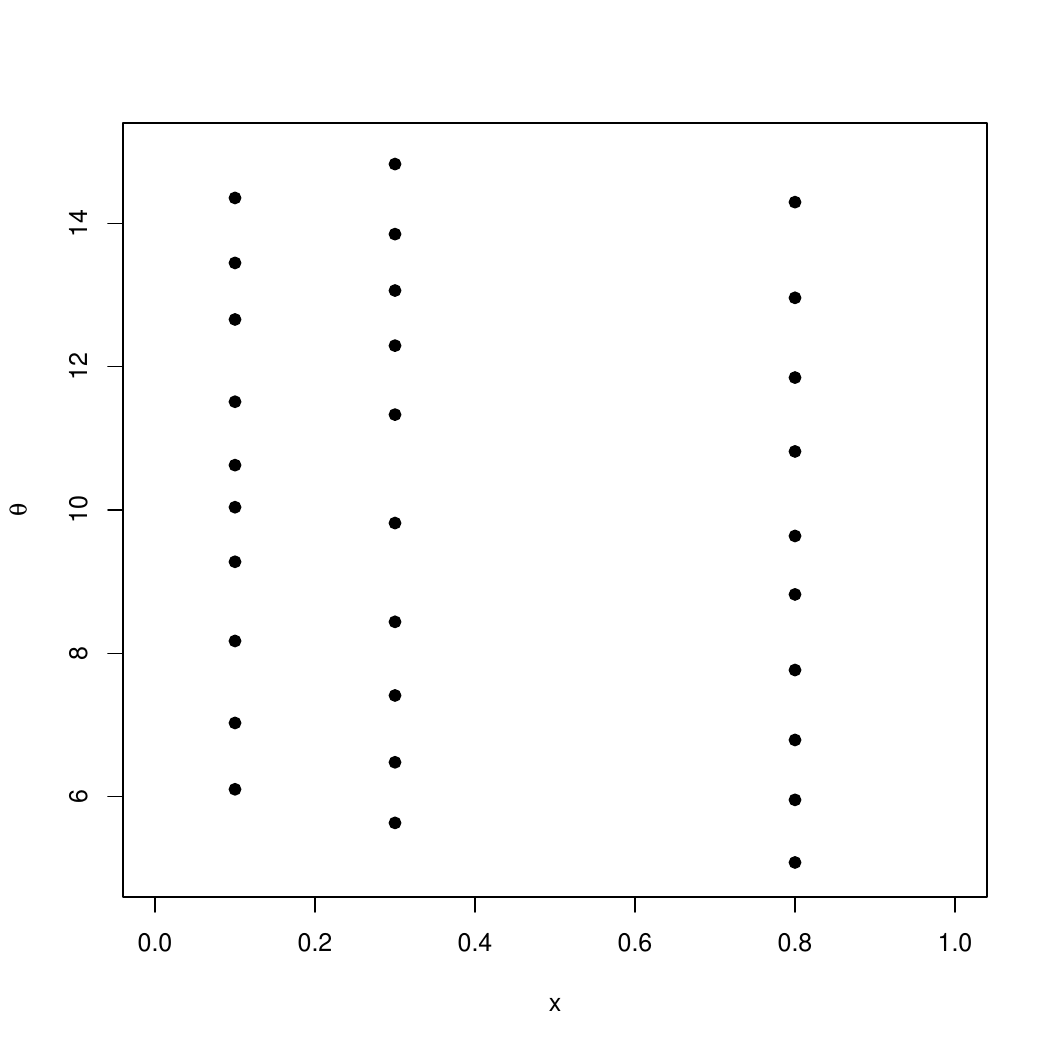}
\includegraphics[scale=0.37]{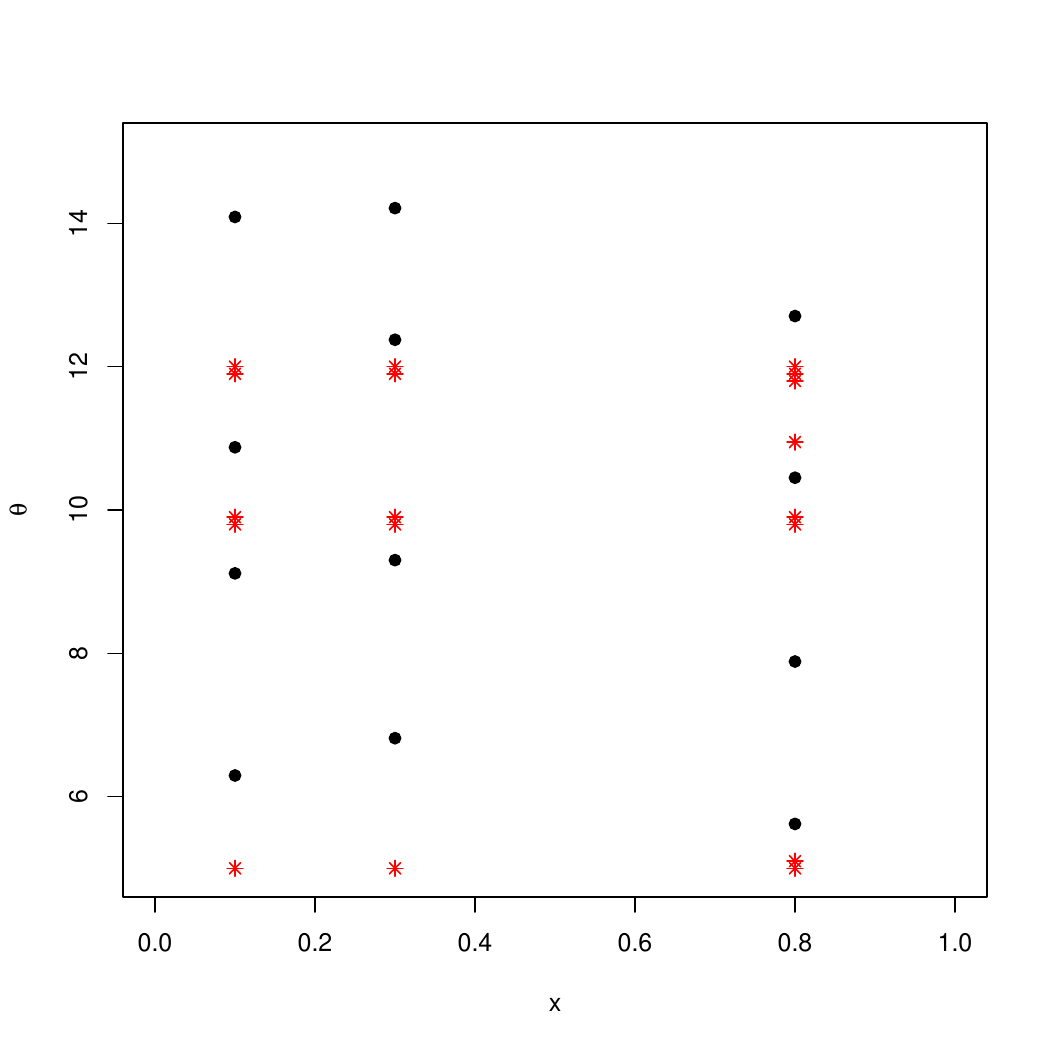}
\label{designs}
\end{figure}

The value of the KL divergence (\ref{KLdiv}) is then computed according to each kind
of design $\mb{D}_M$ used to construct the GPE, namely one shot maximin LHD, one shot restricted-to-$\mb{X}^{f}$ maximin LHD, designs in the \citet{Prat13} fashion and designs constructed with Strategies A, B and C. Let us sum up all the kinds of design of experiments (DOE) $\mb{D}_M$ we try out:

\begin{enumerate}[1.]
\item one-shot maximin LHD on $\mathcal{X}\times\mathcal{T}$. Such a design is displayed in Figure \ref{designs} (upper left),
\item one-shot restricted-to-$\mb{X}^{f}$ maximin LHD. Such a design is displayed in Figure \ref{designs} (bottom left),
\item sequential design obtained with the method of \citet{Prat13} with $6$ iterations of EI,
\item Strategy $A$ starting from a maximin LHD. Such a design is displayed in Figure \ref{designs} (upper right),
\item Strategy $A$ starting from a restricted-to-$\mb{X}^{f}$ maximin LHD. Such a design is displayed in Figure \ref{designs} (bottom right),
\item Strategy $B$ starting from a maximin LHD,
\item Strategy $B$ starting from a restricted-to-$\mb{X}^{f}$ maximin LHD,
\item Strategy $C$ starting from a maximin LHD,
\item Strategy $C$ starting from restricted-to-$\mb{X}^{f}$ maximin LHD.
\end{enumerate} 

\begin{figure}[H]
\centering
\caption{\textit{$\mb{X}^{f}=(0.1,0.3,0.8)^{T}$ Left: boxplots of the KL divergence computed between the APD and the SPD (using the R library FNN). Right: zoom on designs $2$, $3$, $5$, $7$, $9$ }}
\includegraphics[scale=0.38]{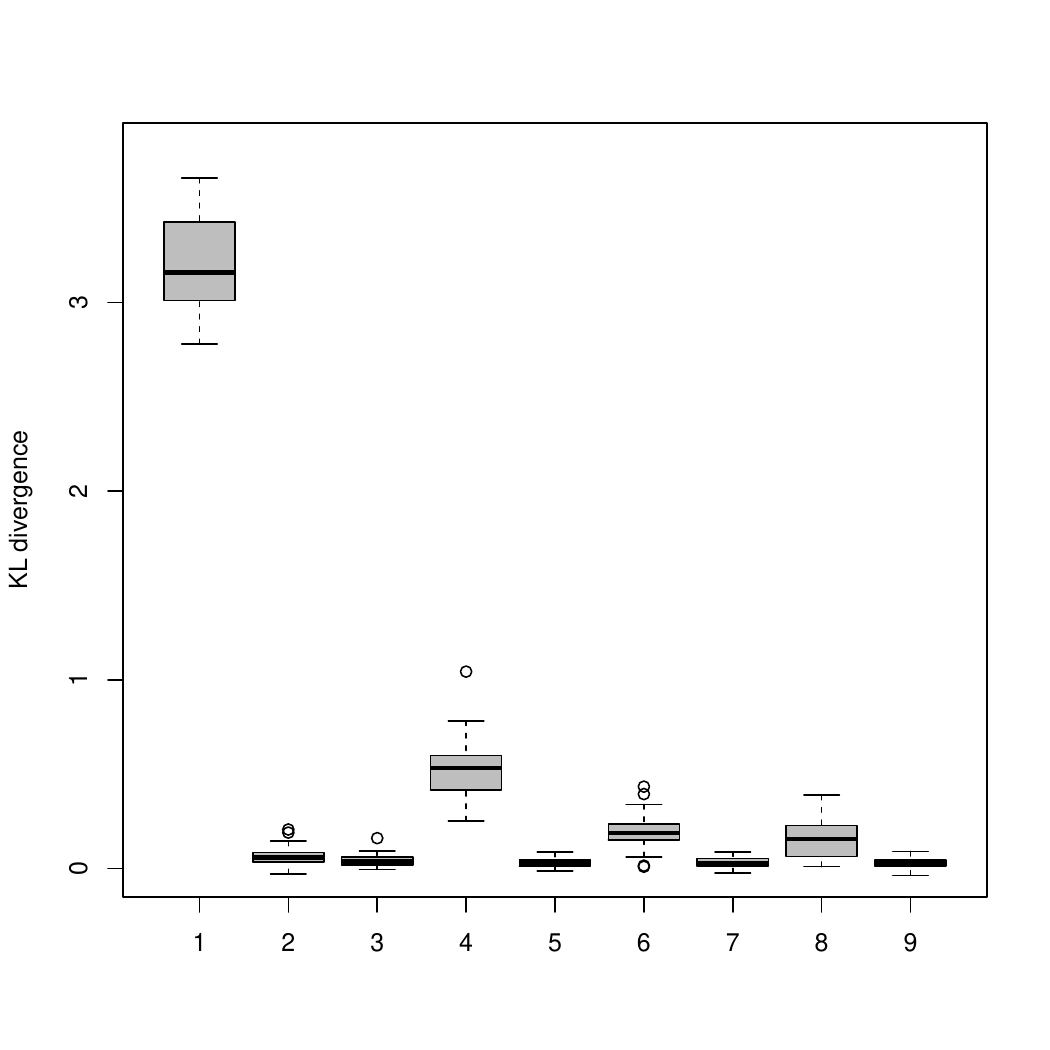}
\includegraphics[scale=0.38]{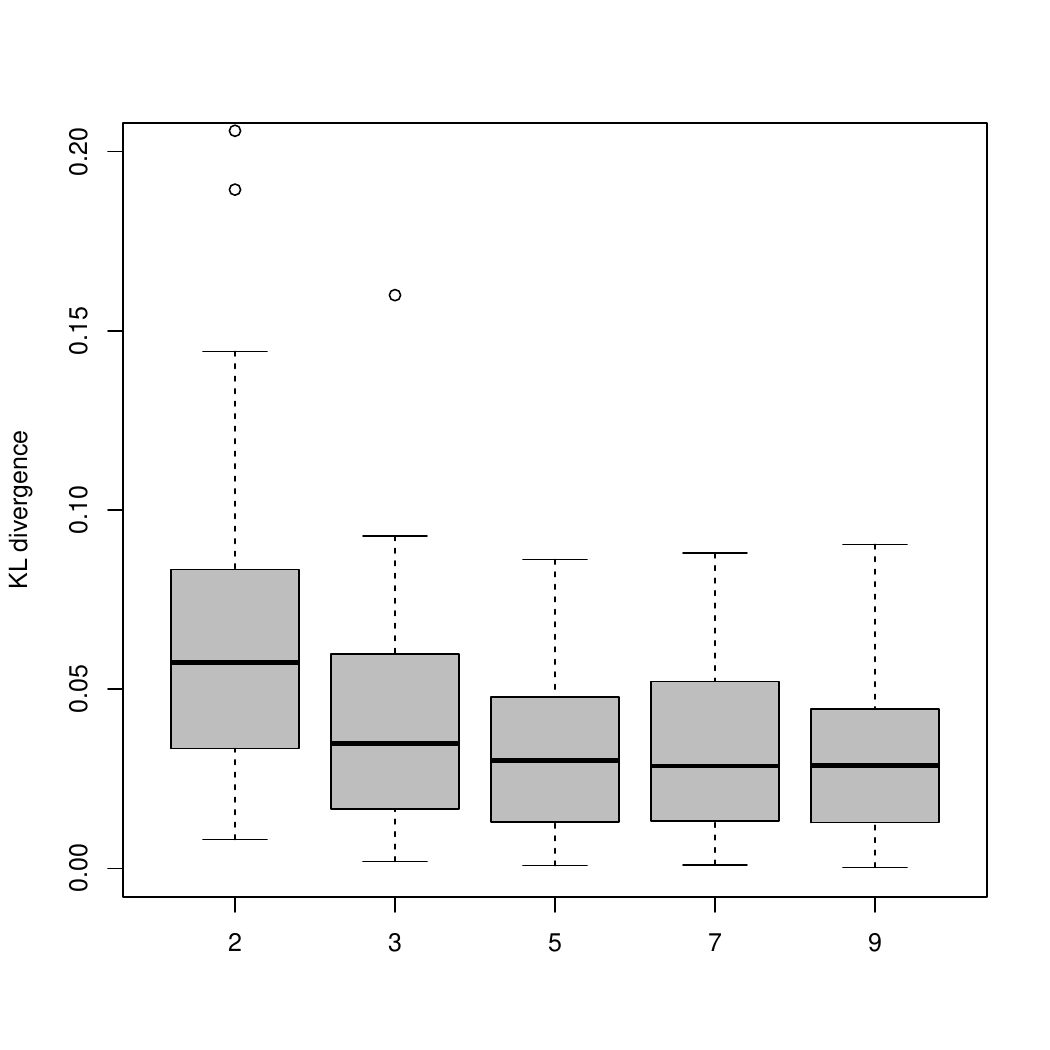}
\label{distributions}
\end{figure}

The robustness of the results is assessed by simulating $50$ different data set $\mb{z}^{f}$ corresponding to $50$ values of $\ms{\theta}$ randomly sampled in $[5,15]$. For each $\mb{z}^{f}$, the SPD is sampled $50$ times according to a different GPE each time. The KL values are averaged over the repeated design generation. 
Figure \ref{distributions} then displays boxplots of the $50$ mean KL divergences for the $9$ different design generation strategies. 
Several observations can be made concerning the results:
\begin{itemize}
\item sequential strategies starting from a maximin LHD outperform one-shot maximin LHD,
\item restricted-to-$\mb{X}^{f}$ maximin LHD outperforms one-shot maximin LHD,
\item one-shot restricted-to-$\mb{X}^{f}$ maximin LHD are as efficient as sequential strategies starting from a restricted-to-$\mb{X}^{f}$ maximin LHD.
\end{itemize}

This first case shows a great interest in constructing a design limited to $\mb{X}^{f}\times\mathcal{T}$ because $30$ locations space filling in $\mb{X}^{f}\times\mathcal{T}$ are enough 
to construct a GPE that quasi perfectly matches the code in this subspace. We are now interested in calibration done with respect to $$\mb{X}^{f}=(0.1,0.2,0.3,0.4,0.5,0.6,0.7,0.8,0.9).$$ 
This second case will show a greater interest than previously in favor of constructing the sequential designs we propose.

\begin{figure}
\centering
\caption{\textit{$\mb{X}^{f}=(0.1,0.2,0.3,0.4,0.5,0.6,0.7,0.8,0.9)^{T}$. Left: the actual posterior distribution. Right: boxplots of the KL divergence computed between the APD and the SPD (using the R library FNN)}}
\includegraphics[scale=0.42]
{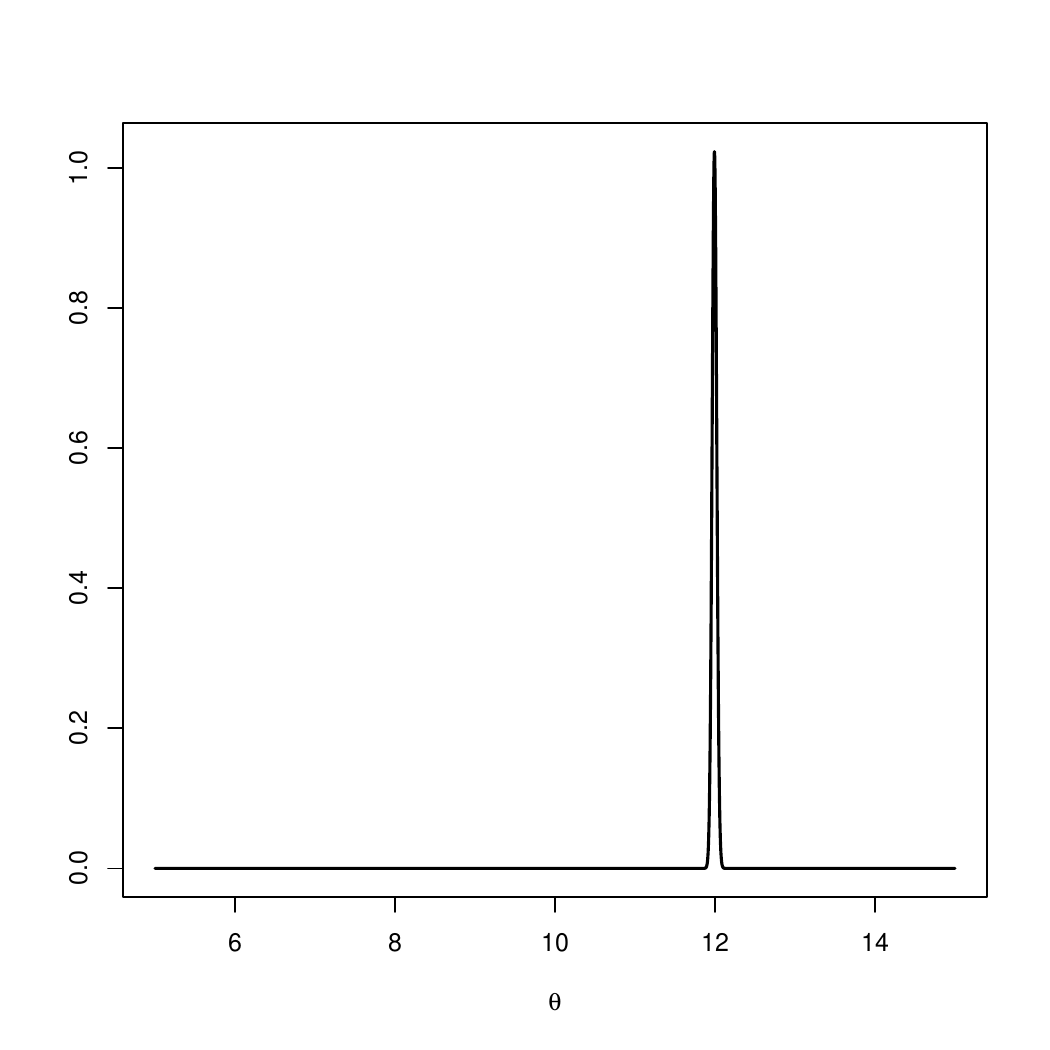}
\includegraphics[scale=0.42]{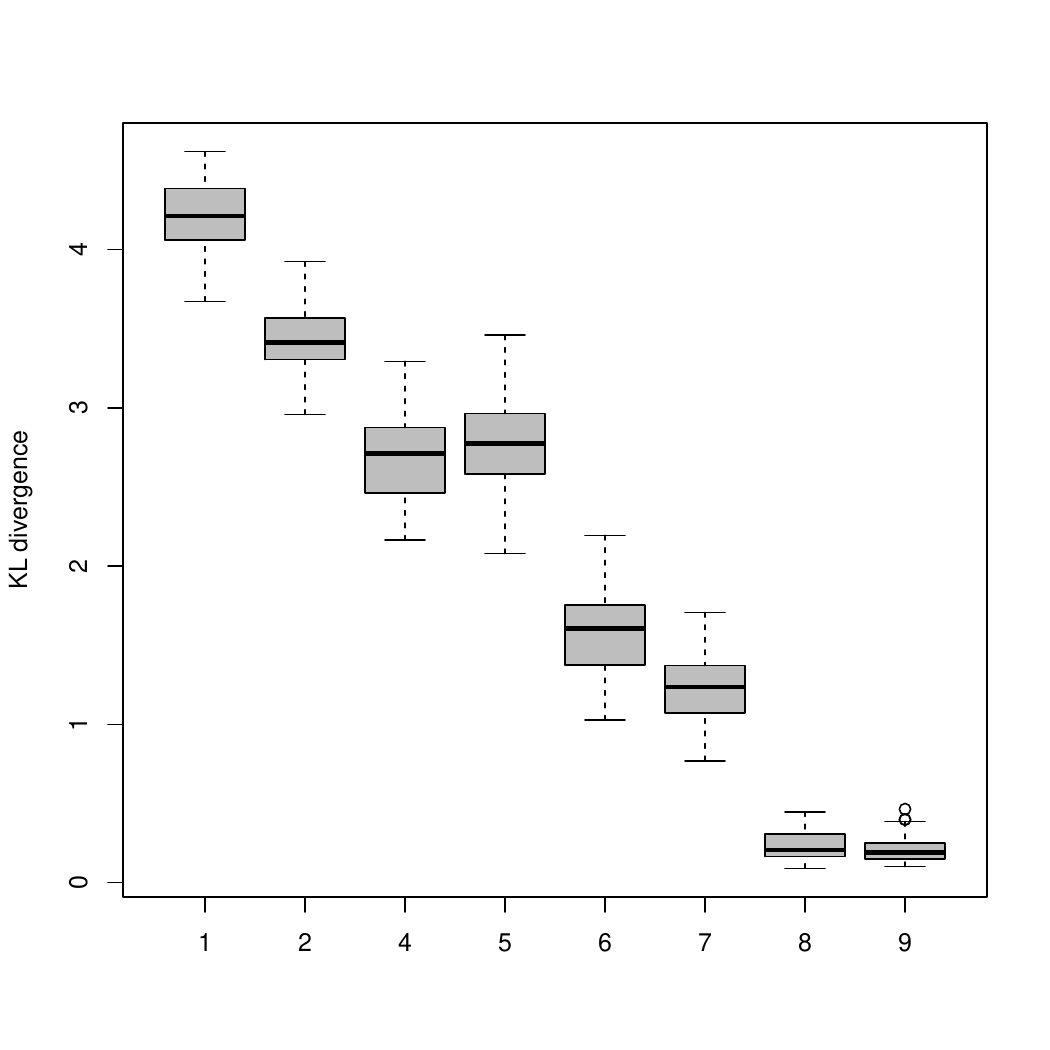}
\label{distributions9}
\end{figure}

\paragraph*{Case $2$: $\mb{X}^f=(0.1,0.2,0.3,0.4,0.5,0.6,0.7,0.8,0.9)^{T}$} The field data $\mb{z}^f$ are still simulated by
\begin{equation}
\label{data2}
z^f_{i}=y_{\theta}(x^f_{i})+\epsilon_i\,\,\,\,\,\textrm{for}\,\,\,\,\,i=1,\cdots,n\,,
\end{equation}
where $\theta=12$ and $\epsilon_i\thicksim\mathcal{N}(0,0.3^2)$.
As $n$ is larger, the APD $\pi(\theta|\mb{z}^f)$ 
now has a single narrow mode around the true value (see Figure \ref{distributions9}, left).
{The SPD 
(\ref{Bayescalib}) is still constructed with a GPE having a constant mean  $m_{\ms{\beta}}=m$ and a Matérn $5/2$ correlation function with $N=30$ simulations}.
As before, the robustness of KL divergence is assessed for Designs $1$ to $9$ except for Design $3$ (which is not expected to perform better 
than Strategy A according to Case $1$ and the discussion after Remark \ref{RemOAT} in Section \ref{sec:three_calib}). 

\begin{figure}
 \centering
\caption{\textit{Case $2$. Upper left: Design $1$. Upper right: Design $4$. Bottom left: Design $2$. Bottom right: Design $9$. 
The black dots are the initial design. The red stars are the new runs selected from the EI criterion.}}
\includegraphics[scale=0.37]{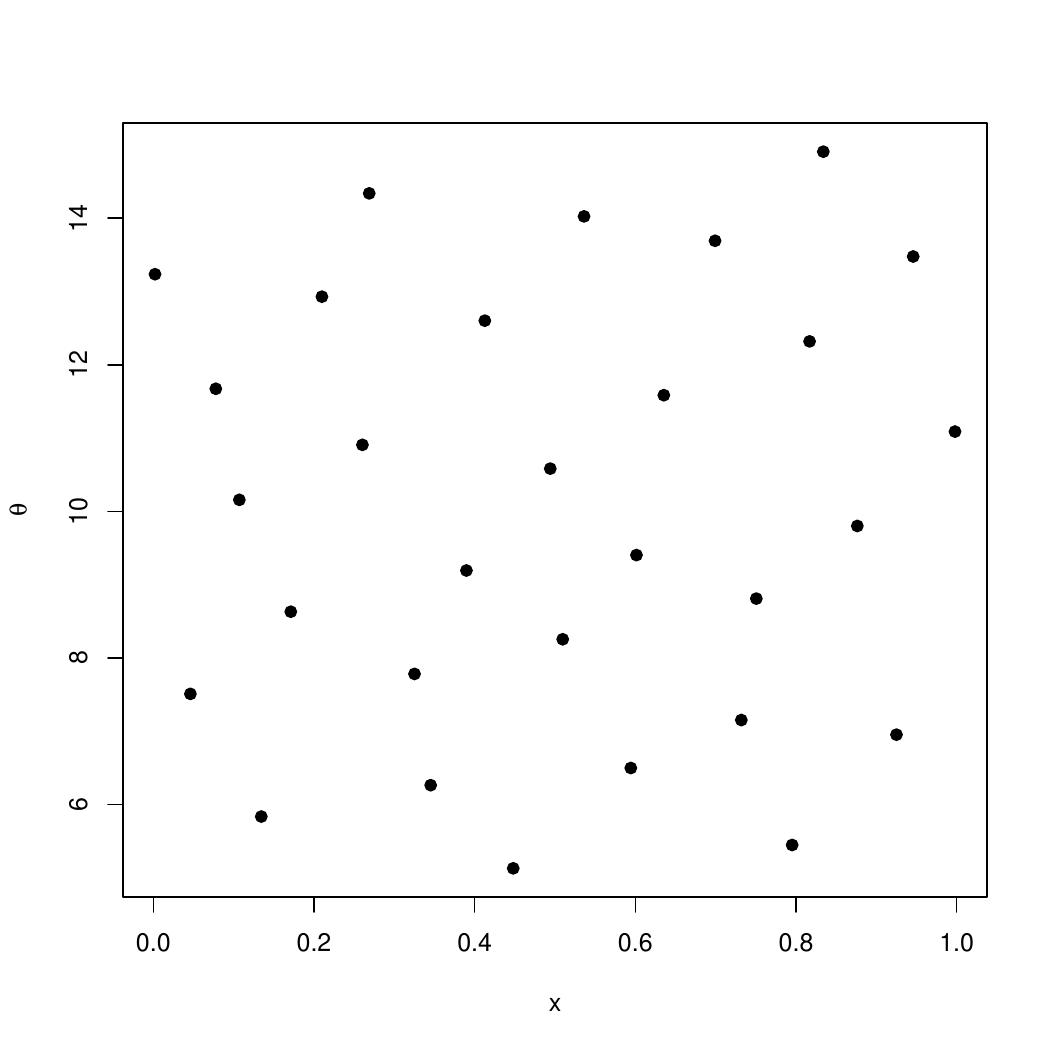}
\includegraphics[scale=0.37]{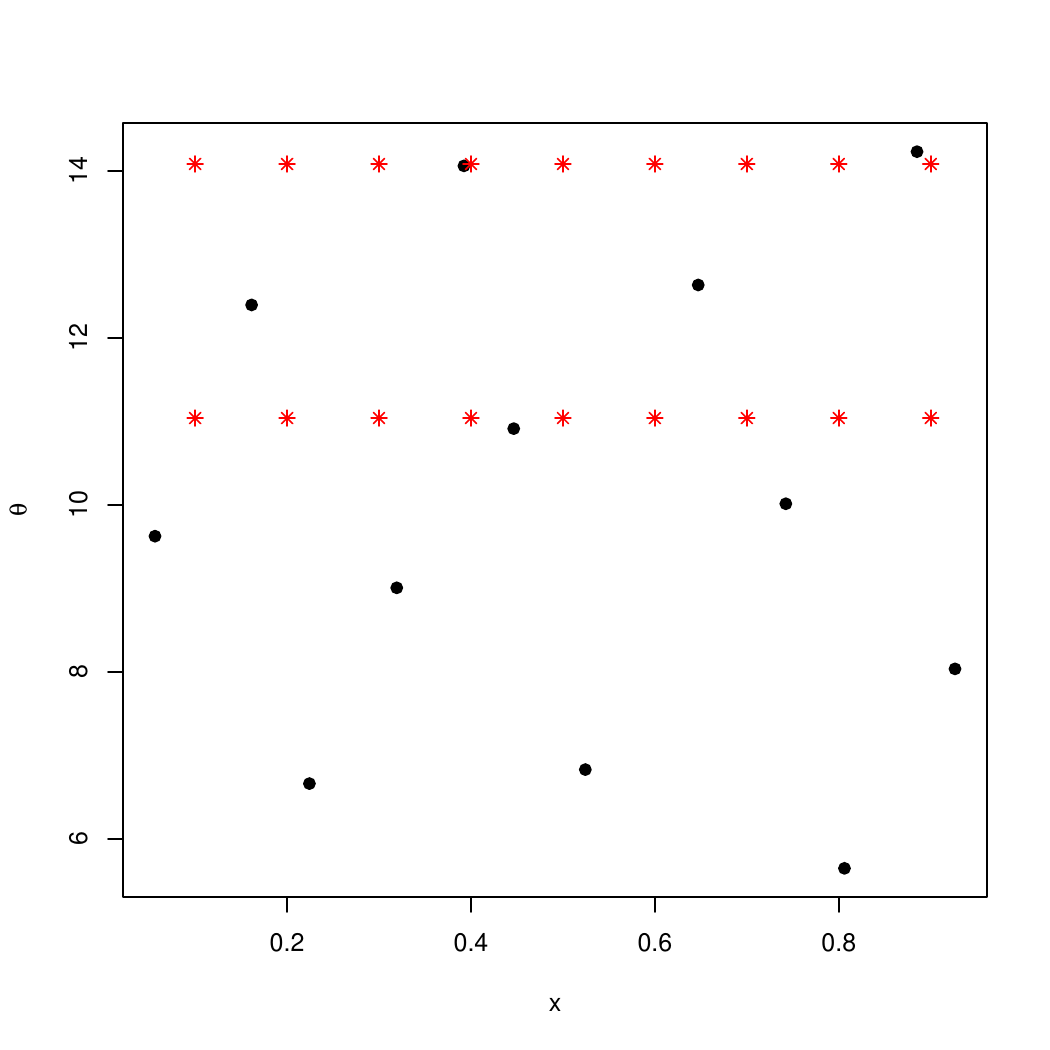}

\includegraphics[scale=0.37]{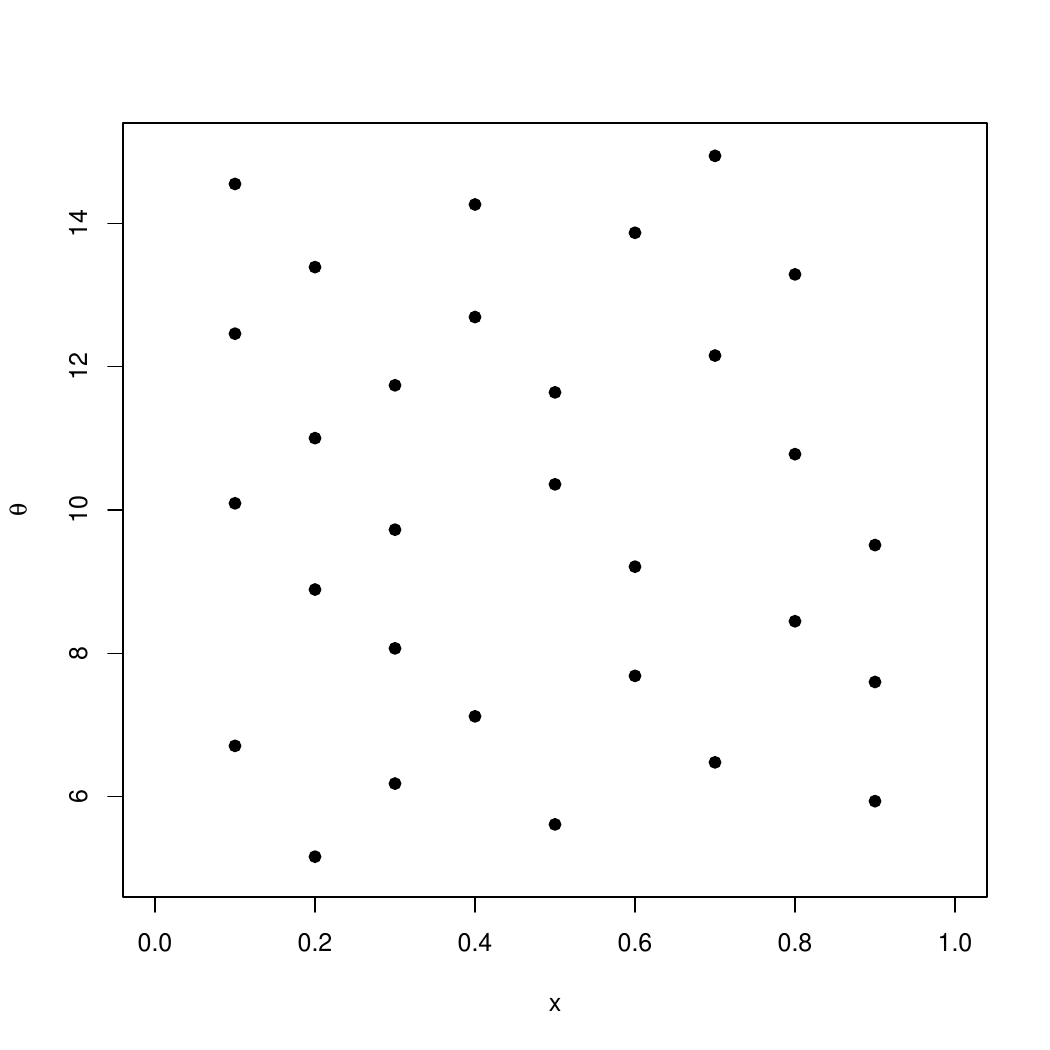}
\includegraphics[scale=0.37]{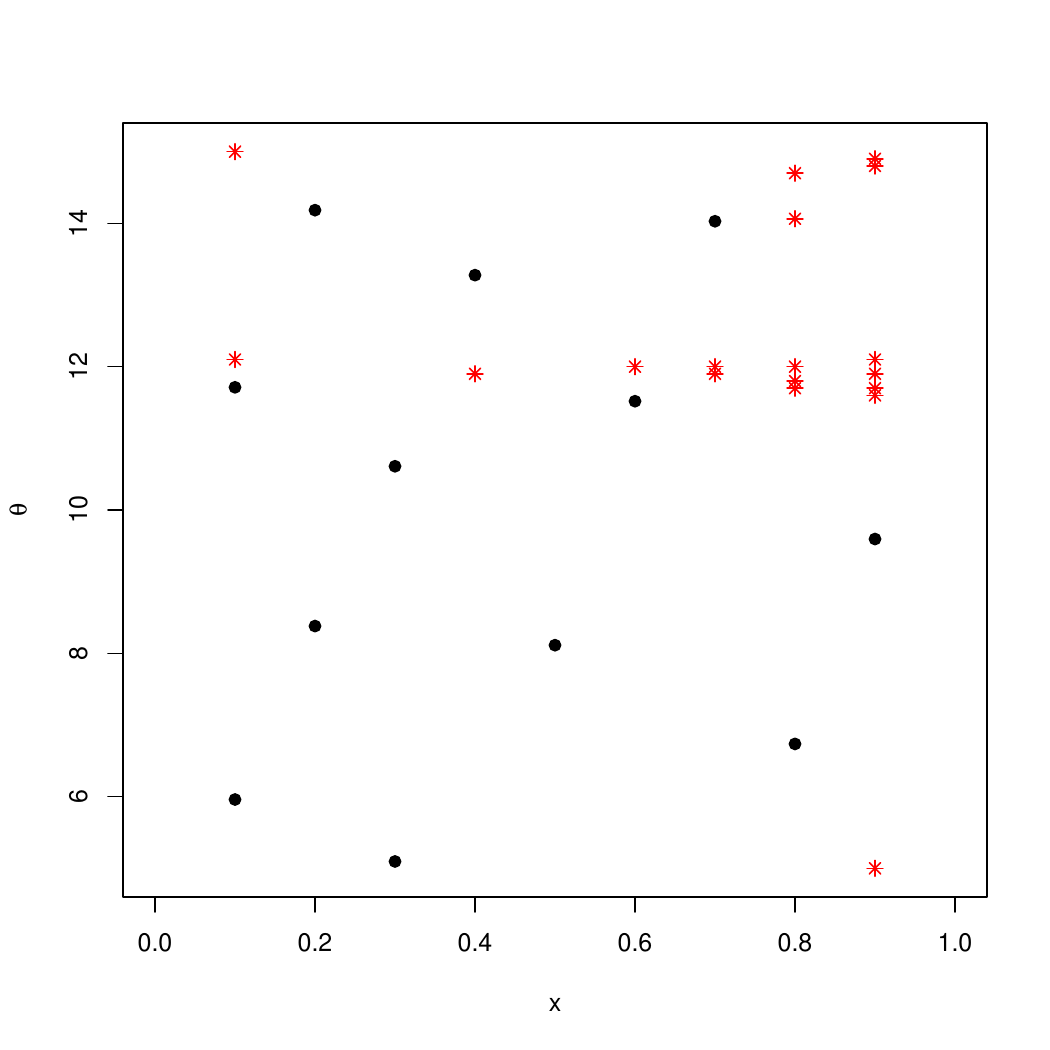}
\label{designs9}
\end{figure}

\medskip
By looking at results in Figure \ref{distributions9} (right) where boxplots are constructed similarly to Case $1$, a couple of informative comments can be done:
\begin{itemize}
\item Whether or not the initial design is restricted to $\mb{X}^{f}$ does not impact the results as much as in the first case.
Figure \ref{designs9} (left) shows both restricted and not restricted designs look quite similar, because in fact $\mb{X}^{f}$ is larger than before and is uniformly sampled along $\mathcal{X}=[0,1]$.
\item Both one-at-a-time Strategy $B$ and $C$ outperform Strategy $A$ because they do not need all the code evaluations around $\theta=14$ to discard this area, making that the GPE is better fitted around the true value $\theta=12$ (see Figure \ref{designs9}, bottom right). 
\item Strategy $C$ appears to work the best. 
\end{itemize}

\paragraph*{A $6$D example}
Let us address a second academic example where the dimension of the control variable $\mb{x}\in\mathcal{X}=[0,1]^{3}$ as well as the dimension of the uncertain parameter $\ms{\theta}\in\mathcal{T}=[0,1]^{3}$ is equal to $3$. It puts in light the interest in higher dimension of using restricted-to-$\mb{X}^{f}$ maximin LHD and sequential designs for GPE-based calibration. The g-Sobol function \citep{Salt00} in 6D plays the role of the computer code. It is written as
\begin{equation}
\label{sobol}
y_{\ms{\tau}}:\mb{x}\in\mathcal{X}\longrightarrow
y_{\ms{\tau}}(\mb{x})=\prod_{i=1}^{3}\frac{|4x_i-2|+\tau_i}{1+\tau_i}\,.
\end{equation}
This highly non-linear function is used within the field of computer experiments to assess the performance of global sensitivity methods \citep{Marrel09, Kuch11}.
The field measurements $\mb{z}^f$ are simulated according to a maximin LHD on $\mathcal{X}$ of size $n=60$. For $1\leq i\leq 60$, we have:
\begin{equation}
z^f_i=y_{\ms{\theta}}(x^f_i)+\epsilon_i\,,
\end{equation}
where ${\epsilon_i}\overset{i.i.d.}{\thicksim}
\mathcal{N}(0,0.05^{2})$ and $\ms{\theta}=(0.55,0.55,0.1)$.
In the same way as the 2D study, we aim at reducing the KL divergence between the SPD (\ref{Bayescalib}) and the APD (\ref{eq4}) where the GPE is fitted with a constant mean $m_{\ms{\beta}}=m$ and a Matérn $5/2$ correlation function. 
Let the prior distribution $\pi(\ms{\theta})$ be uniform on $\mathcal{T}$:
\begin{equation}
\pi({\ms{\theta}})\propto \mb{1}_{[0,1]^{3}}({\ms{\theta}}).
\end{equation}

\begin{figure}[H]
\centering
\caption{$M=150$ : \textit{The solid line represents the actual marginal posterior distribution. The dashed line represents the surrogate marginal posterior distribution using a maximin LHD. The dotted line represents the surrogate marginal posterior distribution using a restricted-to-$\mb{X}^{f}$ LHD. The two-dashed line represents the surrogate marginal posterior distribution using a sequential design.}}
\includegraphics[scale=0.4]{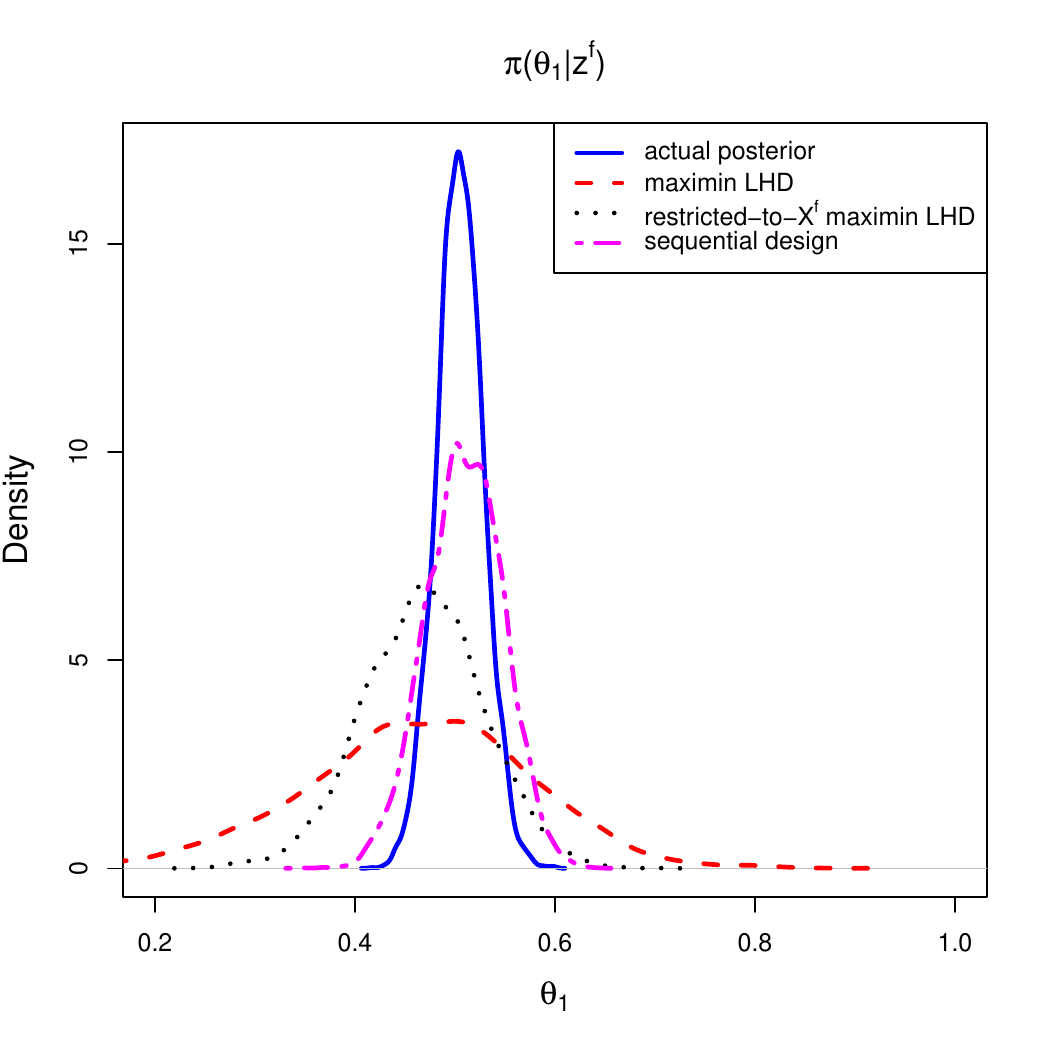}
\includegraphics[scale=0.4]{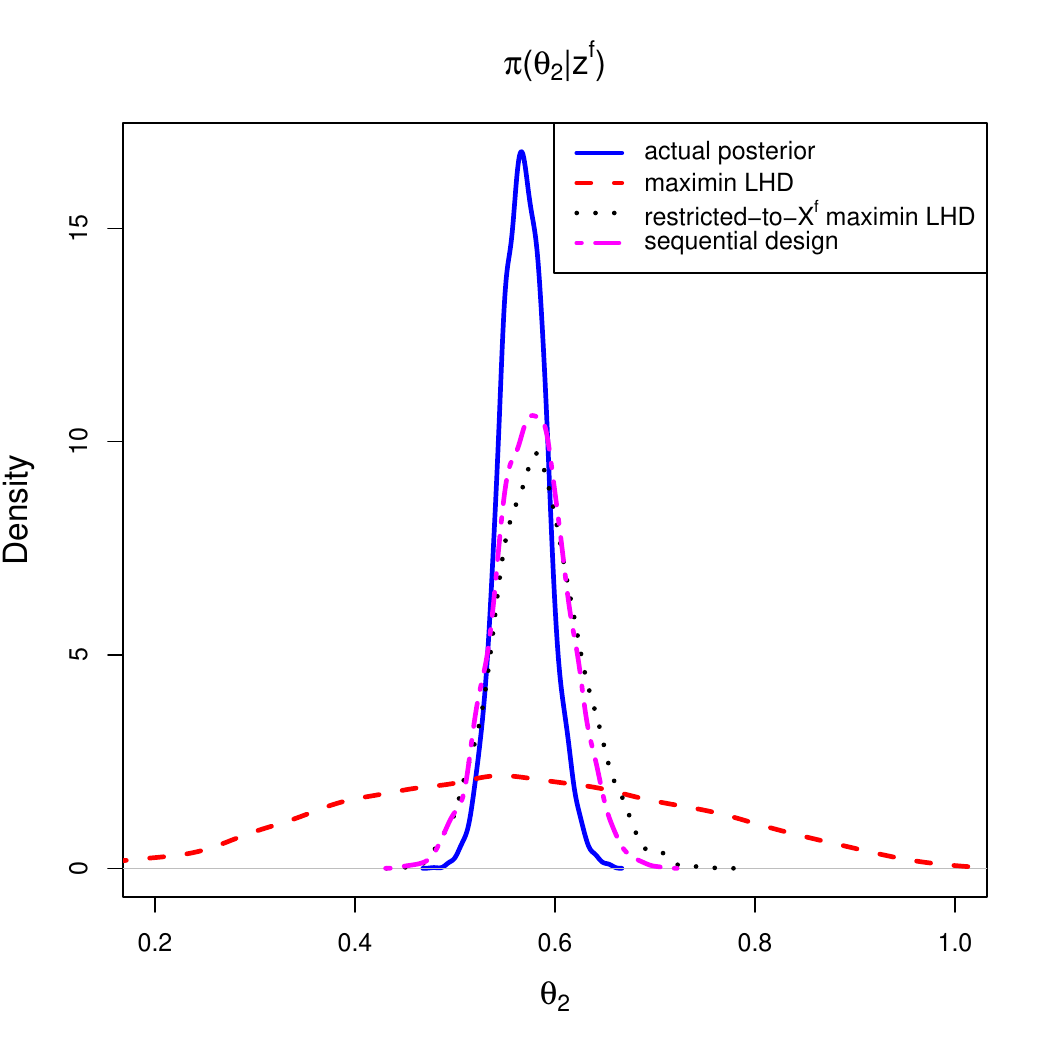}
\includegraphics[scale=0.4]{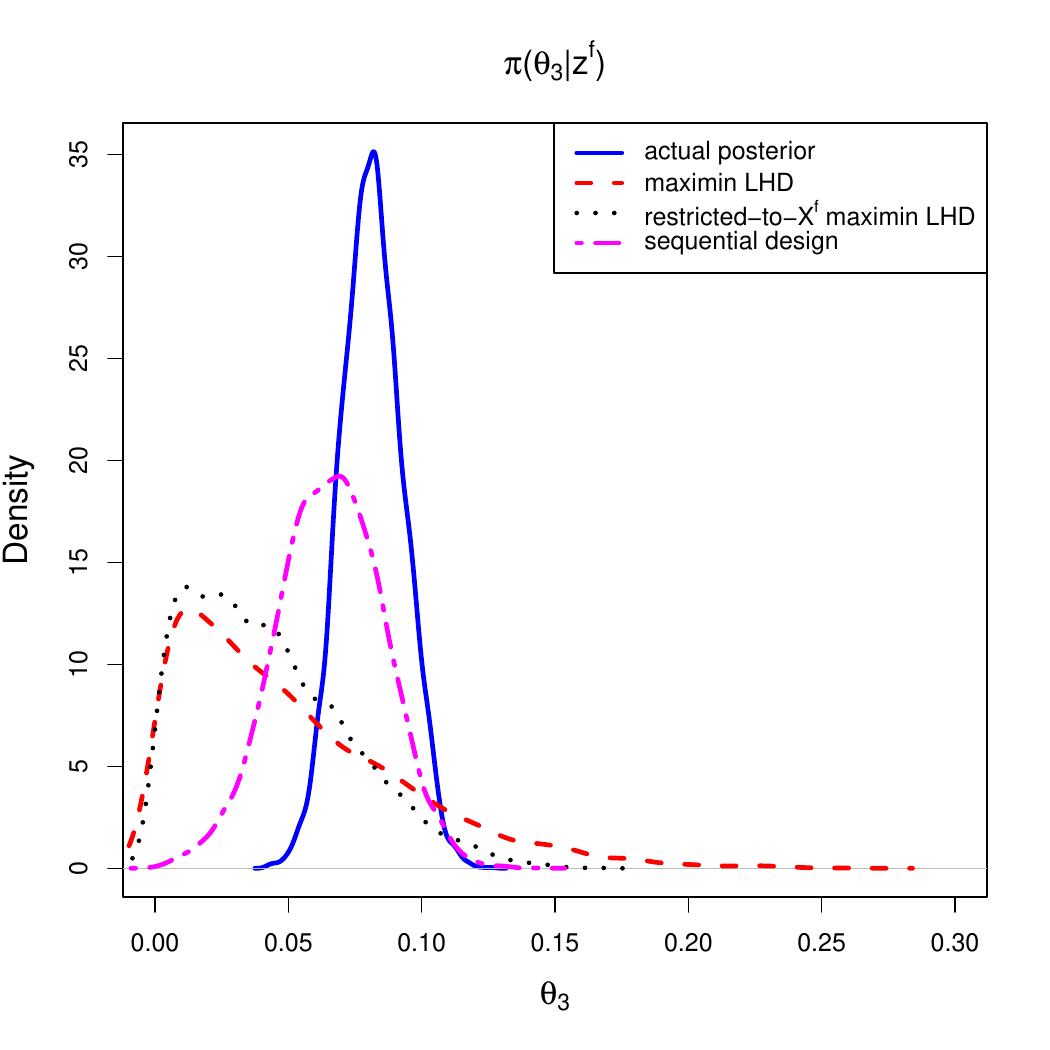}
\label{size150}
\end{figure}

\begin{figure}[H]
\centering
\caption{$M=225$ : \textit{The solid line represents the actual marginal posterior distribution. The dashed line represents the surrogate marginal posterior distribution using a maximin LHD. The dotted line represents the surrogate marginal posterior distribution using a restricted-to-$\mb{X}^{f}$ LHD. The two-dashed line represents the surrogate marginal posterior distribution using a sequential design.}}
\includegraphics[scale=0.4]{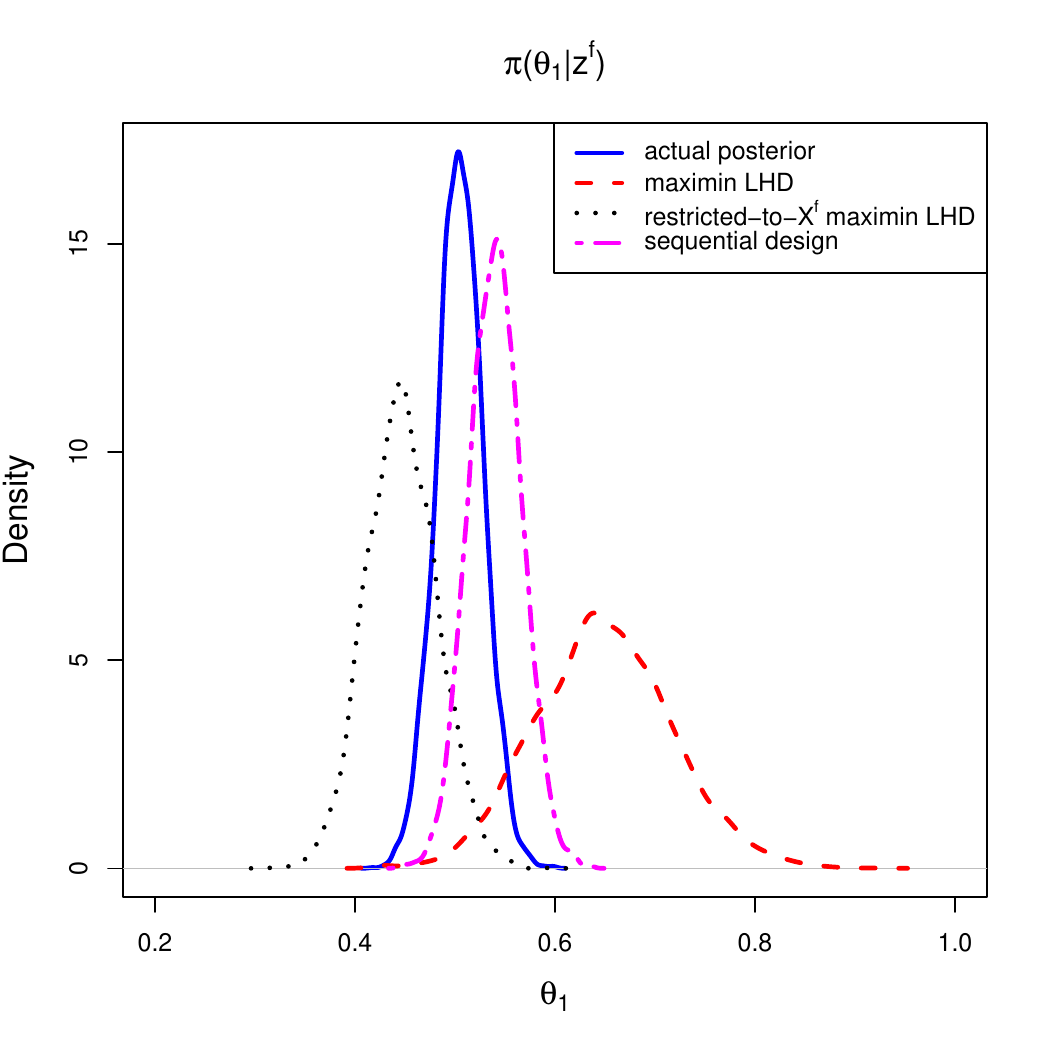}
\includegraphics[scale=0.4]{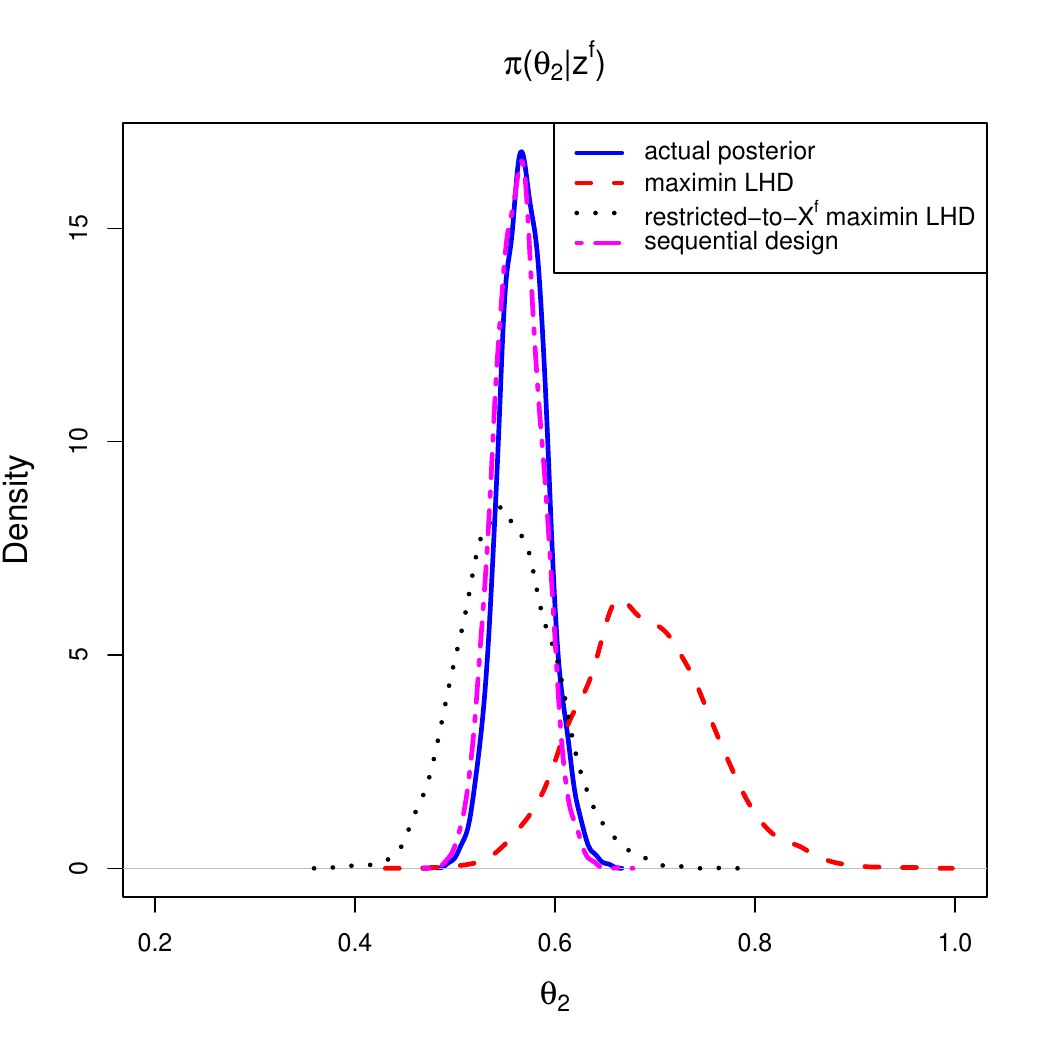}
\includegraphics[scale=0.4]{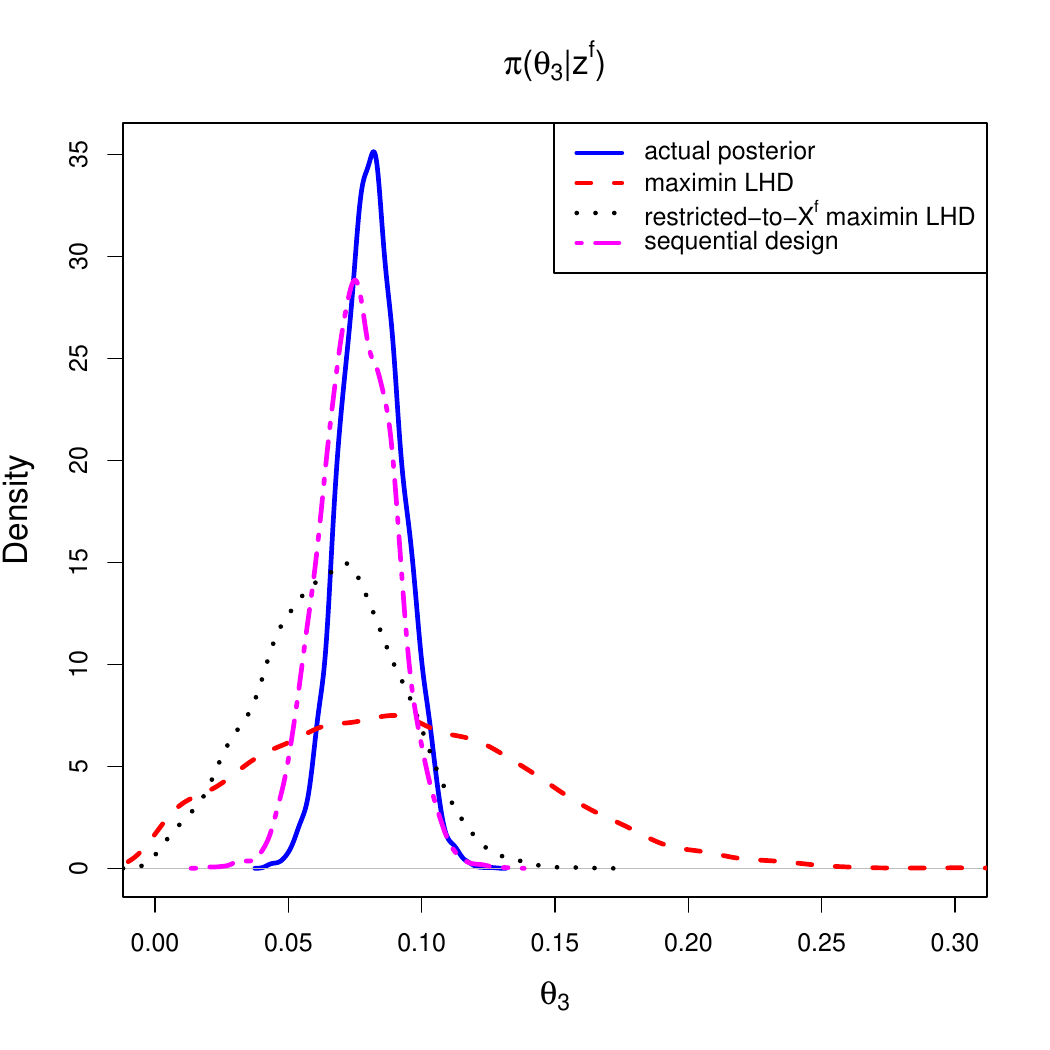}
\label{size225}
\end{figure}

\begin{figure}[H]
\centering
\caption{$M=300$ : \textit{The solid line represents the actual  marginal posterior distribution. The dashed line represents the surrogate marginal posterior distribution using a maximin LHD. The dotted line represents the surrogate marginal posterior distribution using a restricted-to-$\mb{X}^{f}$ LHD. The two-dashed line represents the surrogate marginal posterior distribution using a sequential design.}}
\includegraphics[scale=0.4]{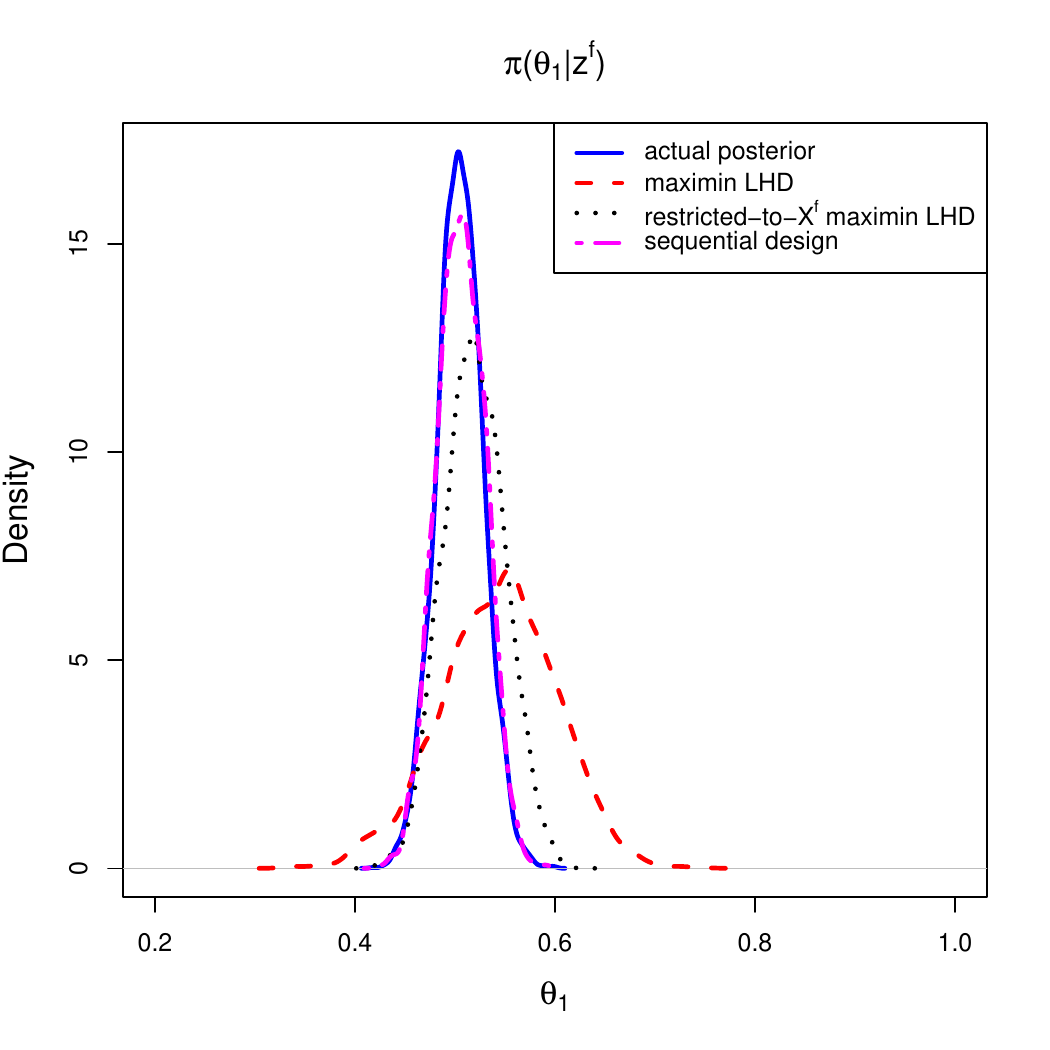}
\includegraphics[scale=0.4]{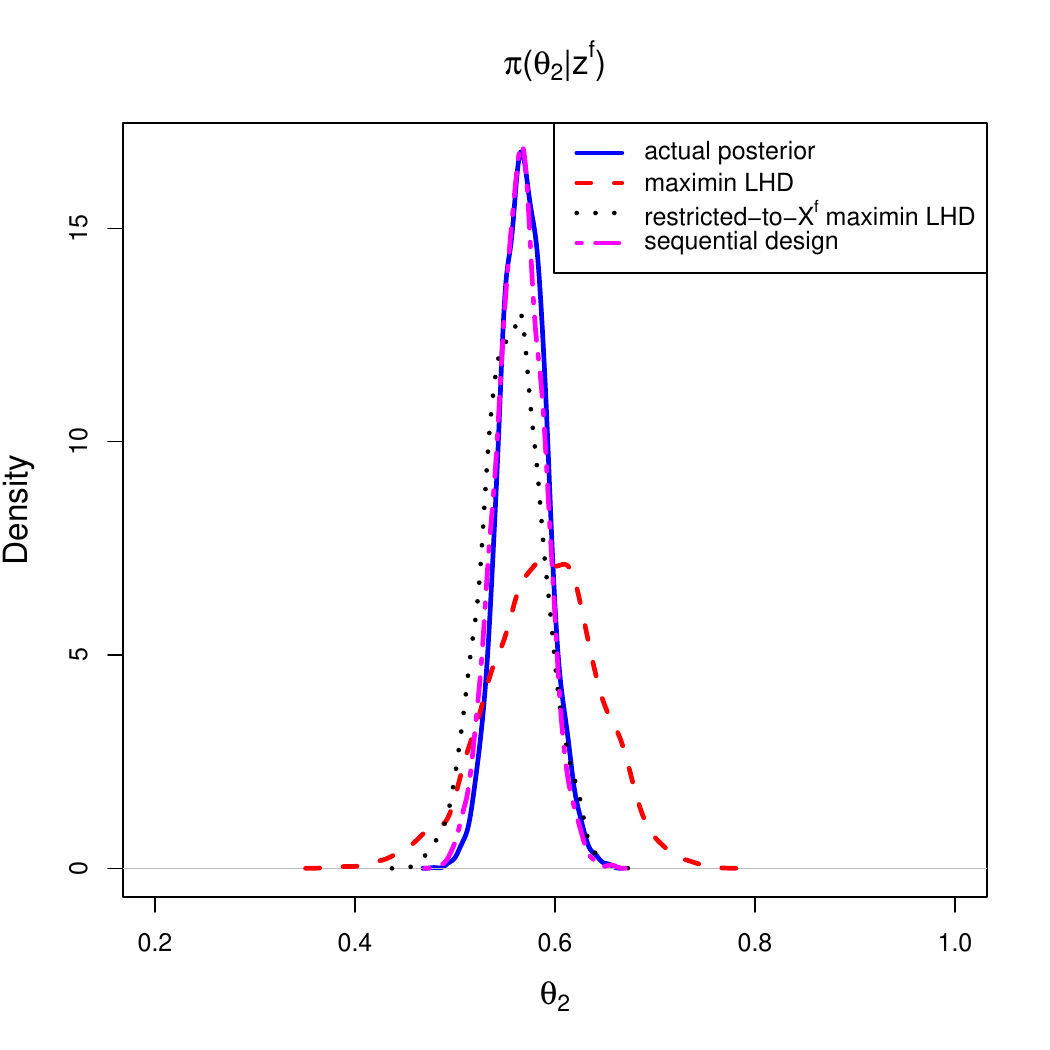}
\includegraphics[scale=0.4]{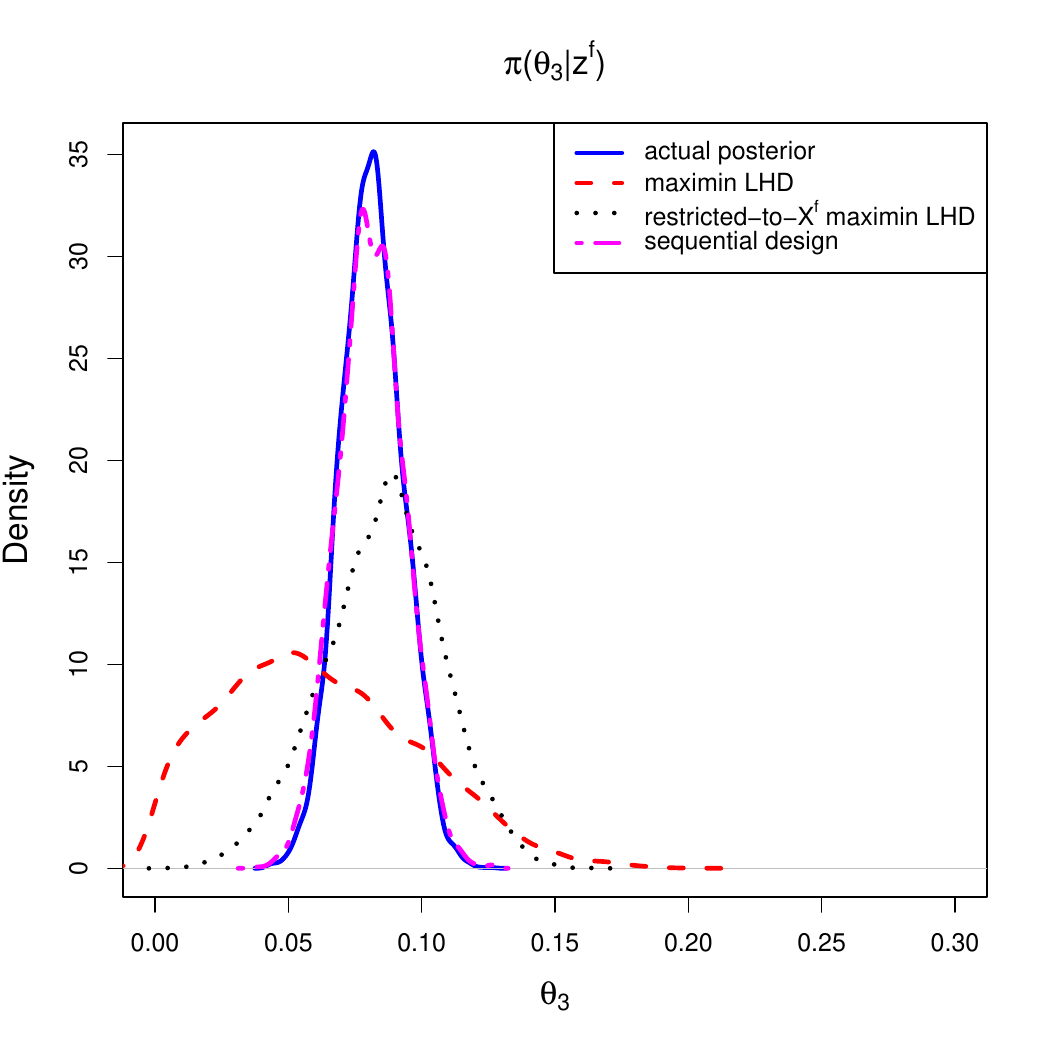}
\label{size300}
\end{figure}

Results are displayed in Figures \ref{size150} to \ref{size300}.
We can see how close the surrogate marginal posterior distributions $\pi(\theta_i|\mb{z}^{f})$ ($i=1,2,3$) are to the actual marginal posterior ones when the sizes of DOEs
are $M=150$, $M=225$ and $M=300$ respectively. In these cases, the default strategy that consists in using a maximin LHD over $\mathcal{X}\times\mathcal{T}$ yields
disappointing results. It is illustrative one more time of the misleading pre-conception that using an
emulator instead of the actual code will result simply in additional uncertainty but in qualitatively
similar results. Although a GPE can deliver good average predictions of the code responses over the input space $\mathcal{X}\times\mathcal{T}$, its uncertainty can be very large within the support of the APD, leading either to a pretty flat posterior or even a strong bias in the actual region of high probability. The use of a restricted-to-$\mb{X}^{f}$ maximin LHD makes a great improvement in terms of minimizing the gap between the SPD and the APD. This gap can be reduced more effectively by using a sequential design constructed with the help of a one-at-a-time EGO algorithm. Indeed, the corresponding surrogate marginal posterior distributions and the actual ones cannot almost be distinguished (see Figure \ref{size300}). The sequential designs have been constructed in a nested way including three consecutive batches of $75$ locations where the initial design was constructed as a restricted-to-$\mb{X}^{f}$ maximin LHD with $M_0=75$ locations.

In the same way as the $2$D study, the research of the location $\ms{\theta}^{\star}$ which maximizes the EGO criterion has been carried out in a greedy fashion. For the first batch, a coarse grid has been used. 
For both the second and third batches, we needed to maximize the EI criterion on a finer grid to better explore the actual regions of high probability.

\begin{figure}[H]
\centering
\caption{\textit{Left: boxplots of the KL divergence computed between the target posterior distribution and the surrogate posterior distribution (using the R library FNN)}}
\includegraphics[scale=0.42]{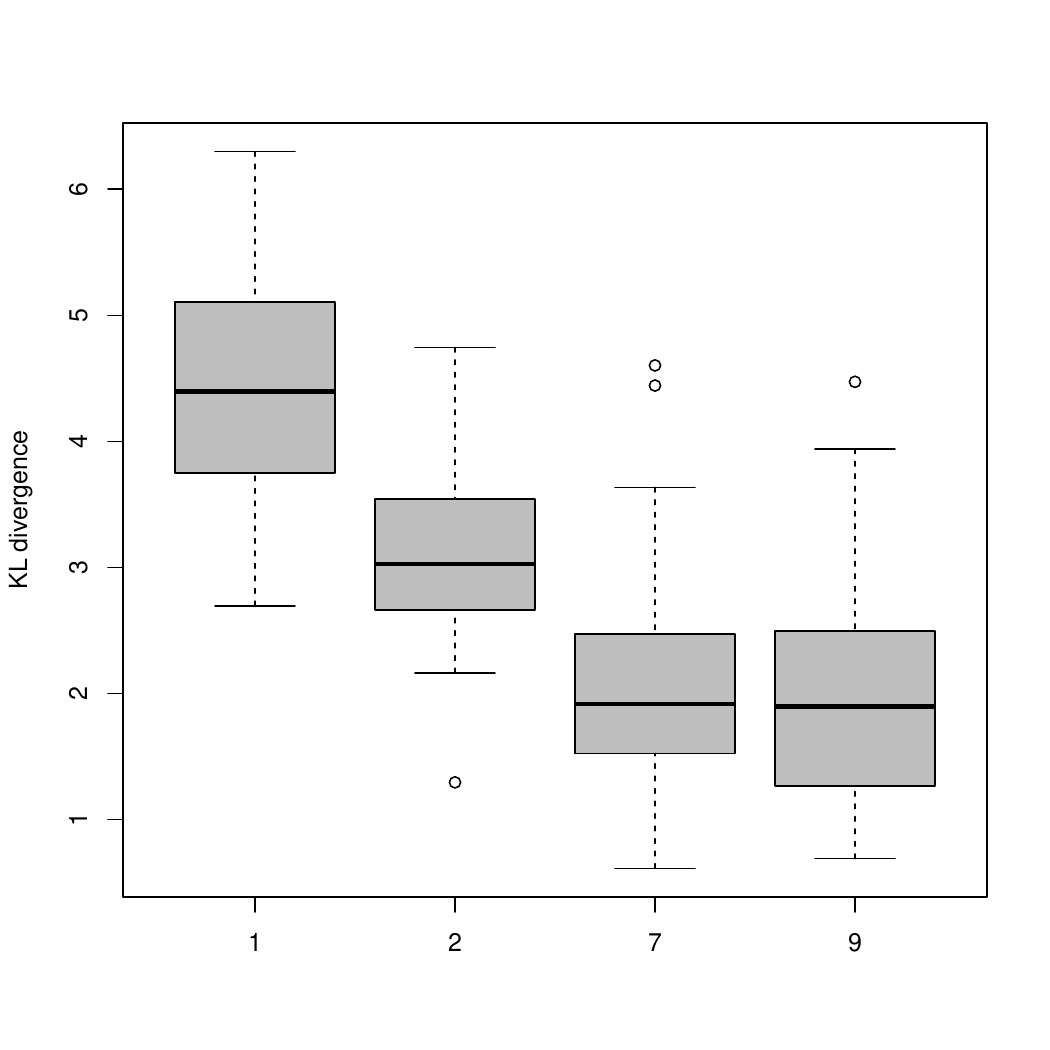}
\label{KL_6D}
\end{figure}

\medskip
We have assessed the robustness of GPE-based calibration by simulating $50$ different data sets $\mb{z}^{f}$ according to $50$ values of $\ms{\theta}$ randomly sampled in $[0,1]^{3}$. For each data set $\mb{z}^{f}$, the surrogate posterior distribution is  sampled $10$ times according to a different GPE each time constructed with $M=150$ simulations. The performance of Designs $1$, $2$, $7$ and $9$ are evaluated in terms of the KL divergence between the SPD and the APD. Results are displayed in Figure \ref{KL_6D} where the boxplots were made with $50$ values (one per data set), each of them being calculated as the mean of the $10$ KL divergence values.
For Designs $7$ and $9$, the initial design $\mb{D}_0$ was constructed as a restricted-to-$\mb{X}^{f}$ LHD of size $M=75$ locations. As the dimension of $\ms{\theta}$ is larger than in the 2D study, more attention has been paid to the grid $G$ for optimizing the EI criterion. If $G$ is too coarse, 
some promising area of the parameter space could be actually not explored whereas if $G$ is too fine the computation
time will be drastically increased. As a compromise, we have conducted optimization alternatively on four coarse grids $G_1$, $G_2$, $G_3$, $G_4$ as EGO iterations occur. We have chosen $G_1=[0,0.2,0.4,0.6,0.8,1]^{3}$, $G_2=[0.1,0.3,0.5,0.7,0.9]^{3}$, $G_3=[0.05,0.25,0.45,0.65,0.85]^{3}$ and $G_4=[0.15,0.35,0.55,0.75,0.95]^{3}$ where
\begin{equation}
G_1\cap G_2\cap\cdots\cap G_N = \varnothing\,.
\end{equation}
Results are displayed in Figure \ref{KL_6D}. They still support  
the advantage of using a restricted-to-$\mb{X}^{f}$ design constructed in an adaptive fashion. They can however appear less impressive than in Case $2$ of the 2D study for essentially two reasons:
\begin{itemize}
\item $150$ locations are not enough for drastically reducing  the uncertainty of $SS(\ms{\theta})$ where it is small. Here, one-at-a-time strategies only add sequentially $75$ locations to an initial design while the size of $\mb{z}^{f}$ is $60$;
\item none grid among $G_1,G_2,G_3,G_4$ covers the unknown true value $\ms{\theta}$ (recall that it is randomly sampled in $[0,1]^{3}$). A finer grid decomposition would make $G$ closer to the true $\ms{\theta}$, making the results even better.
\end{itemize}
Figure \ref{size300} has actually shown that using $M=300$ locations instead of $M=150$ to construct sequentially the GPE as well as using a finer grid in the region of high posterior density is likely to lead to a perfect agreement between the SPD and the APD.

\medskip
Lastly, Strategy $B$ and Strategy $C$ yield similar results because all the field locations in $\mb{X}^{f}$ have comparable impact on the variation of (\ref{sobol}) with respect to $\ms{\tau}$.
In such a case, both Criteria (\ref{firstcrit}) and (\ref{second_crit_exact}) select $\mb{x}^{\star}$ in a rather close way that is based on the variance of the GPE.



\section{Conclusion}

\label{sec:five_calib}

This paper deals with new adaptive numerical designs for calibrating time-consuming computer codes in a Bayesian setting. 
For such codes, Bayesian calibration is based on a Gaussian process emulator (GPE) which approximates the code output and thus makes the MCMC algorithms practicable. After choosing a design of experiments, the GPE is estimated by using a modular approach which allows us to separate the estimation of GPE parameters from the estimation of the code parameters.

Our contribution consists in taking advantage of the stochastic property of the GPE to 
construct sequentially the design of experiments in such a way that the gap between the posterior distribution based on the GPE (the so-called surrogate posterior distribution) and the posterior distribution based on the code (the so-called actual posterior distribution) is the smallest possible in terms of the KL divergence. We have shown that it is of a great importance to reduce the uncertainty of the GPE where the density of the actual posterior distribution is large. In an objective Bayesian context where no prior expertise is available about the unknown parameters, 
this goal is equivalent to reduce the uncertainty of the GPE where the sum of squares of the residuals between the code outputs and the field measurements is small.
We have thus proposed sequential strategies for constructing the design of experiments based on the EI criterion designed for the sum of squares of the residuals. Our simulations on academic examples have shown that such designs outperform space-filling designs with respect
to the KL divergence. \NNNew{These sequential strategies must be initialized from a first design of experiments which can be chosen to be space-filling.}
However, simulations have been performed in an unbiased framework where there is no discrepancy between the physical system and the computer code. If prior information is available about the shape of the code discrepancy, our algorithms could be applied to a weighted sum of squares function in a similar way.


It may appear surprising that the prior distribution is not taken into account in the EI maximization \NNew{since it could in fact be taken into account without extra difficulty.}
{This choice can be defended for two main reasons. In the one hand, the GPE that is constructed from an adaptive design does not depend on the prior distribution and thus we are able to perform several calibrations under different prior distributions as well as a \NNew{sensitivity} analysis to the prior choice. In the other hand, if the prior distribution is locally uniform (as vague prior distributions are) then minimizing the sum 
of squares of the residuals is quite close to maximizing the posterior distribution. 
Our method could also be applied using ``non-informative" priors such as Jeffreys prior or Berger-Bernardo prior because they intend to have a negligible influence in the face of data. In practice, {unfortunately}, such priors can lead to improper posteriors and be hard to compute for a costly computer code, making them difficult to use in the context of this work.}


{The main interest of the EGO algorithms presented in the paper is that the objective function, that is the sum of squares
of the residuals, is not itself modeled by a Gaussian process, which makes it possible to conduct either an exact EGO or a version of it that is one-at-a-time.
The latter consists in adding sequentially a single input location \NNew{which is a couple} of a new parameter value and a value of the control variable restricted to the field locations, to the current design by maximizing the EI criterion. However, a search over the whole space might be a better option in some cases. The initial design was constructed as either a maximin LHD 
or a maximin LHD restricted to the field locations. 
The reported simulation studies have shown that the second class of designs works better although we think a maximin LHD could be more relevant if new fields measurements were collected during the calibration process. Besides, EGO algorithms should be designed according to the computing resources in order to run as many new simulations as available nodes.}

{Another concern is how well the EI criterion is maximized over the parameter space as the iterations of the algorithm occur. It is clear that the effectiveness of the optimization technique determines the efficiency of the sequential strategies (in particular when the dimension of the parameter is high). Because in the method, the EI criterion has no closed form expression, its optimization has been crudely performed in a greedy fashion. In the future, free-derivative optimization algorithms should be tested \citep{Conn09}, perhaps improving the results.}
\NNNew{Eventually, the refinement of the approximation of the code within a Metropolis-Hastings algorithm proposed by \citet{conrad2016} is another
sequential procedure which incorporates new runs where the posterior distribution is high. Although the local approximation can be based on 
local GPE, their algorithm differs from the usual Bayesian calibration algorithms. A comparison of the two methods is an interesting perspective 
for future works.}

\section*{Acknowledgments}
This work was partially funded by the French Agence Nationale pour la Recherche (ANR), under grant ANR-13-MONU-0005 (project CHORUS).
The authors want to thank Joan Sobota for English proofreading.
The authors also thank the Associate Editor and the two reviewers for their very valuable comments {which have led to a significant improvement of the paper.}

\section*{Proof of Proposition \ref{PropKL}}

The log-conditional likelihood is referred to as $\ell^{C}(\ms{\theta};\mb{z}^f|y(\mb{D}_M))=\log\big(\mathcal{L}^C(\ms{\theta};\mb{z}^f|y(\mb{D}_M))\big)$
and the target log-likelihood is referred to as $\ell(\ms{\theta};\mb{z}^f)=\log\big(\mathcal{L}((\ms{\theta};\mb{z}^f)\big)$.
It is sufficient to prove that
\begin{equation}
|\ell^{C}(\ms{\theta};\mb{z}^f|y(\mb{D}_M))-\ell(\ms{\theta};\mb{z}^f)|
\label{log_diff} 
\end{equation}
is uniformly bounded in $\ms{\theta}$ and the bound tends to zero with $M\rightarrow \infty$.
We can decompose (\ref{log_diff}) as
\begin{multline}
\label{twotermsKL}
|\ell^{C}(\ms{\theta};\mb{z}^f|y(\mb{D}_M))-\ell(\ms{\theta};\mb{z}^f)|\le
|\ell^{C}(\ms{\theta};\mb{z}^f|y(\mb{D}_M))-\tilde \ell^{C}(\ms{\theta};\mb{z}^f|y(\mb{D}_M))|+\\
|\tilde \ell^{C}(\ms{\theta};\mb{z}^f|y(\mb{D}_M))-\ell(\ms{\theta};\mb{z}^f)|
\end{multline}
where 
\begin{equation*}
\tilde \ell^{C}(\ms{\theta};\mb{z}^f|y(\mb{D}_M))=-\frac{1}{2\lambda^2}
(\mb{z}^f-
\mu^{M}_{\ms{\beta},\ms{\Psi}}(\mb{D}_{\ms{\theta}}))^{\textrm{T}}
(\mb{z}^f-\mu^{M}_{\ms{\beta},\ms{\Psi}}(\mb{D}_{\ms{\theta}}))
-\frac{n}{2}\log{{2\pi}}-n\log{\lambda}
\end{equation*}
corresponds to the log-conditional likelihood
where the function $y$ is replaced with $\mu^{M}_{\ms{\beta},\ms{\Psi}}$ and the 
covariance matrix of the GPE is neglected.

\medskip
The second term is bounded as:
\begin{eqnarray*}
|\tilde \ell^{C}(\ms{\theta};\mb{z}^f|y(\mb{D}_M))-\ell(\ms{\theta};\mb{z}^f)|&=&
\bigg|-\frac{1}{2\lambda^2}\bigg(
||\mb{z}^f-
\mu^{M}_{\ms{\beta},\ms{\Psi}}(\mb{D}_{\ms{\theta}})||^2
-||\mb{z}^f-y_{\ms{\theta}}(\mb{X}^f)||^{2}
\bigg)\bigg|\\
&=&\bigg|-\frac{1}{2\lambda^2}\bigg(\sum_{i=1}^n(y_{\ms{\theta}}(\mb{x}^f_{i})-
\mu^{M}_{\ms{\beta},\ms{\Psi}}\big(\mb{x}^f_{i},\ms{\theta})\big)\big(2z^f_{i}-y_{\ms{\theta}}(\mb{x}^f_{i})-\mu^{M}_{\ms{\beta},\ms{\Psi}}(\mb{x}^f_{i},\ms{\theta})\big)\bigg|\,.
\end{eqnarray*}
Let us suppose that the minimax distance, say $(h_{\mb{D}_M})_M$, of the designs sequence $(\mb{D}_M)_M$ tends to $0$, namely
\begin{equation}
h_{\mb{D}_M}=\max_{(\mb{x}',\ms{\tau}')\in\mathcal{X}\times\mathcal{T}}\min_{(\mb{x}^D_i,\ms{\tau}^D_i)\in \mb{D}_M}\Vert 
(\mb{x}',\ms{\tau}')-(\mb{x}_i,\ms{\tau}_i)\Vert\underset{M\to\infty}\longrightarrow 0
\end{equation}
Then, the uniform bound is deduced from the point-wise bound 
given for standard radial basis correlation function ${C_{\ms{\psi}}}$ \citep{Scha95b}. We can obtain
\begin{equation}
 |y_{\ms{\theta}}(\mb{x}^f_{i})-
\mu^{M}_{\ms{\beta},\ms{\Psi}}(\mb{x}^f_{i},\ms{\theta})|\le \Vert y \Vert_{C_{\ms{\psi}}}\cdot G_{C_{\ms{\psi}}}(h_{\mb{D}_M})\,,
\end{equation}
where
$\Vert y \Vert_{C_{\ms{\psi}}}$ is the norm of $y$ in the RKHS associated to ${C_{\ms{\psi}}}$ and $G_{C_{\ms{\psi}}}(.)$ tends to $0$ when $h_{\mb{D}_M}$ tends to $0$.

\medskip
Using the triangle inequality, an upper bound for the first term in (\ref{twotermsKL}) is written as,
\begin{multline}
 |\ell^{C}(\ms{\theta};\mb{z}^f|y(\mb{D}_M))-\tilde \ell^{C}(\ms{\theta};\mb{z}^f|y(\mb{D}_M))|\leq 
 \Big|\frac{1}{2}\log(|V^{M}_{\ms{\Psi},\sigma^2}(\ms{\theta})+\lambda^2\mb{I}_n|)-\frac{n}{2}\log(\lambda^2)\Big| \\
 +\Big|\frac{1}{2\lambda^2}\Vert\mb{z}^f-
\mu^{M}_{\ms{\beta},\ms{\Psi}}(\mb{D}_{\ms{\theta}})\Vert^2  
-\frac{1}{2}
(\mb{z}^f-
\mu^{M}_{\ms{\beta},\ms{\Psi}}(\mb{D}_{\ms{\theta}}))^{\textrm{T}}(V^{M}_{\ms{\Psi},\sigma^2}(\ms{\theta})+\lambda^2\mb{I}_n)^{-1}
(\mb{z}^f-\mu^{M}_{\ms{\beta},\ms{\Psi}}(\mb{D}_{\ms{\theta}}))\Big|
\label{Scha2}
\end{multline}
Then,
\begin{eqnarray}
 \Big|\frac{1}{2}\log(|V^{M}_{\ms{\Psi},\sigma^2}(\ms{\theta})+\lambda^2\mb{I}_n|)-\frac{n}{2}\log(\lambda^2)\Big| 
 \nonumber
 &=&\bigg|\frac{1}{2}\log\Big(\frac{|V^{M}_{\ms{\Psi},\sigma^2}(\ms{\theta})+\lambda^2\mb{I}_n|}{(\lambda^{2})^{n}}\Big)\bigg|\\
\nonumber 
 &\le&\bigg|\frac{n}{2}\log\Bigg( \frac{\sum_{i=1}^{n}\big(V^{M}_{\ms{\Psi},\sigma^2}(\ms{\theta})_{ii}/n\big)+\lambda^2}{\lambda^2}\Bigg)\bigg|\\
 \nonumber
 &=&\bigg|\frac{n}{2}\log\Big(\frac{\sum_{i}V^{M}_{\ms{\Psi},\sigma^2}(\ms{\theta})_{ii}}{\lambda^2n}+1\Big)\bigg|\\
 \label{Scha1}
\end{eqnarray}
where the inequality of arithmetic and geometric means is used for bounding the determinant of the matrix $V^{M}_{\ms{\Psi},\sigma^2}(\ms{\theta})+\lambda^2\mb{I}_n$ by a function of the trace. Using again a result in \citet{Scha95b}, we have $V^{M}_{\ms{\Psi},\sigma^2}(\ms{\theta})_{_{ii}}\le  G_{C_{\ms{\psi}}}(h_{\mb{D}_M})$. It follows that,
\begin{equation*}
\bigg|\frac{n}{2}\log\Big(\frac{\sum_{i}V^{M}_{\ms{\Psi},\sigma^2}(\ms{\theta})_{ii}}{\lambda^2n}+1\Big)\bigg|\le
 C_y \frac{n}{\lambda^2} G_{C_{\ms{\psi}}}(h_{\mb{D}_M})
\end{equation*}
where $C_y$ is a constant. Since $(h_{\mb{D}_M})_M$ tends to $0$ with $M\to\infty$, (\ref{Scha1}) tends to $0$ with $M\to\infty$.

\smallskip
For the second term (\ref{Scha2}), we use the series expansion $(A+\mb{I}_ n)^{-1}=\mb{I}_n+\sum_{i=1}^{\infty}(-1)^i A^i$, valid when $||A||<1$. It can thus be applied to $V^{M}_{\ms{\Psi},\sigma^2}(\ms{\theta})\lambda^{-2}$ because both inequalities $V^{M}_{\ms{\Psi},\sigma^2}(\ms{\theta})_{ii}\le  G_{C_{\ms{\psi}}}(h_{\mb{D}_M})$ and $|V^{M}_{\ms{\Psi},\sigma^2}(\ms{\theta})_{ij}| \leq \sqrt{V^{M}_{\ms{\Psi},\sigma^2}(\ms{\theta})_{ii}V^{M}_{\ms{\Psi},\sigma^2}(\ms{\theta})_{jj}}$ ensure its norm is lower than $1$ once $M$ is large enough. Therefore, 
\begin{multline*}
\Big|\frac{1}{2\lambda^2}\Vert\mb{z}^f-
\mu^{M}_{\ms{\beta},\ms{\Psi}}(\mb{D}_{\ms{\theta}})\Vert^2  
-\frac{1}{2}
(\mb{z}^f-
\mu^{M}_{\ms{\beta},\ms{\Psi}}(\mb{D}_{\ms{\theta}}))^{\textrm{T}}(V^{M}_{\ms{\Psi},\sigma^2}(\ms{\theta})+\lambda^2\mb{I}_n)^{-1}
(\mb{z}^f-\mu^{M}_{\ms{\beta},\ms{\Psi}}(\mb{D}_{\ms{\theta}}))\Big|= \\
\Big|\frac{1}{2\lambda^2}
(\mb{z}^f-
\mu^{M}_{\ms{\beta},\ms{\Psi}}(\mb{D}_{\ms{\theta}}))^{\textrm{T}}\Big(\sum_{i=1}^{\infty}(-1)^i[V^{M}_{\ms{\Psi},\sigma^2}(\ms{\theta})\lambda^{-2}]^{i}\Big)(\mb{z}^f-
\mu^{M}_{\ms{\beta},\ms{\Psi}}(\mb{D}_{\ms{\theta}}))\Big|
\end{multline*}
which tends to $0$ with $M\to\infty$, that completes the proof of the uniform convergence of (\ref{log_diff}) to $0$ with $M\to\infty$.

\bibliographystyle{apalike}

\begin{thebibliography}{}

\bibitem[Bachoc, 2014]{Bach14}
Bachoc, F. (2014).
\newblock Asymptotic analysis of the role of spatial sampling for covariance
  parameter estimation of {G}aussian processes.
\newblock {\em Journal of Multivariate Analysis}, 125:1--35.

\bibitem[Bastos and O'Hagan, 2008]{Bast08}
Bastos, L. and O'Hagan, A. (2008).
\newblock Diagnostics for {G}aussian process emulators.
\newblock {\em Technometrics}, 51(4):425--438.

\bibitem[Bayarri et~al., 2007]{Bay07}
Bayarri, M., Berger, J., Paulo, R., Sacks, J., Cafeo, J., Cavendish, J., Lin,
  C., and Tu, J. (2007).
\newblock A framework for validation of computer models.
\newblock {\em Technometrics}, 49(2):138--154.

\bibitem[Box and Tiao, 1973]{BoxTiao73}
Box, G. E.~P. and Tiao, G.~T. (1973).
\newblock {\em {B}ayesian Inference in Statistical Analysis}.
\newblock Addison-Wesley, Reading.

\bibitem[Brynjarsdottir and O'Hagan, 2014]{Bry_OHagan14}
Brynjarsdottir, J. and O'Hagan, A. (2014).
\newblock Learning about physical parameters: The importance of model
  discrepancy.
\newblock {\em Inverse Problems}, 30(11):3251--3269.

\bibitem[Campbell, 2006]{Cam06}
Campbell, K. (2006).
\newblock Statistical calibration of computer simulations.
\newblock {\em Reliability Engineering and System Safety},
  91(10-11):1358--1363.

\bibitem[Conn et~al., 2009]{Conn09}
Conn, A., Sheinberg, K., and Vicente, L. (2009).
\newblock {\em Introduction to Derivative-Free Optimization}.
\newblock MOS SIAM Series on Opimization, Reading.

\bibitem[Conrad et~al., 2016]{conrad2016}
Conrad, P.~R., Marzouk, Y.~M., Pillai, N.~S., and Smith, A. (2016).
\newblock Accelerating asymptotically exact {MCMC} for computationally
  intensive models via local approximations.
\newblock {\em Journal of the American Statistical Association},
  111(516):1591--1607.

\bibitem[Cover and Thomas, 1991]{Cover06}
Cover, T. and Thomas, J. (1991).
\newblock {\em Elements of Information Theory}.
\newblock Wiley-Interscience.

\bibitem[Cox et~al., 2001]{Cox01}
Cox, D., Park, J., and Clifford, E. (2001).
\newblock A statistical method for tuning a computer code to a data base.
\newblock {\em Computational Statistics and Data Analysis}, 37(1):77--92.

\bibitem[Craig et~al., 2001]{Craig01}
Craig, P., Goldstein, M., Rougier, J., and Seheult, A. (2001).
\newblock Bayesian forecasting for complex systems using computer simulators.
\newblock {\em Journal of the American Statistical Association},
  96(454):717--729.

\bibitem[Craig et~al., 1997]{Craig97}
Craig, P.~S., Goldstein, M., Seheult, A.~H., and Smith, J.~A. (1997).
\newblock {\em Pressure Matching for Hydrocarbon Reservoirs: A Case Study in
  the Use of Bayes Linear Strategies for Large Computer Experiments}, pages
  37--93.
\newblock Springer, New York, NY.

\bibitem[Currin et~al., 1991]{Currin91}
Currin, C., Mitchell, T., Morris, M., and Ylvisaker, D. (1991).
\newblock Bayesian prediction of deterministic functions, with applications to
  the design and analysis of computer experiments.
\newblock {\em Journal of the American Statistical Association},
  86(416):953--963.

\bibitem[Damblin et~al., 2016]{Dam16}
Damblin, G., Keller, M., Barbillon, P., Pasanisi, A., and Parent, E. (2016).
\newblock Bayesian model selection for the validation of computer codes.
\newblock {\em Quality and Reliability Engineering International},
  32(6):2043--2054.

\bibitem[Ellis and Maitra, 2007]{Ellis12}
Ellis, N. and Maitra, R. (2007).
\newblock Multivariate {G}aussian simulation outside arbitrary ellipsoids.
\newblock {\em Journal of Computational and Graphical Statistics},
  16(3):692--708.

\bibitem[Fang et~al., 2006]{Fang}
Fang, K., Li, R., and Sudijianto, A. (2006).
\newblock {\em Design and modeling for computer experiments}.
\newblock Computer Science and Data Analysis. Chapman \& Hall.

\bibitem[Fang et~al., 2005]{fang:2005}
Fang, K., Li, R., and Sudjianto, A. (2005).
\newblock {\em Design and Modeling for Computer Experiments (Computer Science
  \& Data Analysis)}.
\newblock Chapman \& Hall/CRC.

\bibitem[Gang et~al., 2009]{Gang09}
Gang, H., Santner, T., and Rawlinson, J. (2009).
\newblock Simultaneous determination of tuning and calibration parameters for
  computer experiments.
\newblock {\em Technometrics}, 51(4):464--474.

\bibitem[Ginsbourger, 2009]{Ginsbourger}
Ginsbourger, D. (2009).
\newblock {\em Multiples Métamodèles pour l'approximation et l'optimisation
  de fonctions numériques multivariables}.
\newblock PhD thesis, Ecole Nationale Supérieure des Mines de Saint-Etienne.

\bibitem[Gramacy et~al., 2015]{Gramacy14}
Gramacy, R.~B., Bingham, D., Holloway, J.~P., Grosskopf, M.~J., Kuranz, C.~C.,
  Rutter, E., Trantham, M., and Drake, R.~P. (2015).
\newblock Calibrating a large computer experiment simulating radiative shock
  hydrodynamics.
\newblock {\em Ann. Appl. Stat.}, 9(3):1141--1168.

\bibitem[Higdon et~al., 2004]{Higd04}
Higdon, D., Kennedy, M., Cavendish, J., Cafeo, J., and Ryne, R. (2004).
\newblock Combining field data and computer simulations for calibration and
  prediction.
\newblock {\em SIAM Journal on Scientific Computing}, 26(2):448--466.

\bibitem[Janusevskis and Le~Riche, 2013]{Janis12}
Janusevskis, J. and Le~Riche, R. (2013).
\newblock Simultaneous kriging-based estimation and optimization of mean
  response.
\newblock {\em Journal of Global Optimization}, 55(2):313--336.

\bibitem[Jones et~al., 1998]{Jones98}
Jones, D., Schonlau, M., and Welch, W. (1998).
\newblock Efficient global optimization of expensive black-box functions.
\newblock {\em Journal of Global Optimization}, 13(4):455--492.

\bibitem[Joseph and Melkote, 2009]{Joseph09}
Joseph, V. and Melkote, S. (2009).
\newblock Statistical adjustments to engineering models.
\newblock {\em Journal of Quality Technology}, 41(4):362--375.

\bibitem[Kennedy and O'Hagan, 2001]{Koh2001}
Kennedy, M. and O'Hagan, A. (2001).
\newblock Bayesian calibration of computer models (with discussion).
\newblock {\em Journal of the Royal Statistical Society, Series B,
  Methodological}, 63(3):425--464.

\bibitem[Koehler and Owen, 1996a]{Owen}
Koehler, J. and Owen, A. (1996a).
\newblock Computer experiments.
\newblock {\em In Design and Analysis, Handbook of Statistics, Vol.13}, pages
  261--308.

\bibitem[Koehler and Owen, 1996b]{koehler:owen:1996}
Koehler, J.~R. and Owen, A.~B. (1996b).
\newblock Computer experiments.
\newblock In {\em Design and analysis of experiments}, volume~13 of {\em
  Handbook of Statist.}, pages 261--308. North-Holland, Amsterdam.

\bibitem[Kucherenko et~al., 2011]{Kuch11}
Kucherenko, S., Feil, B., Shah, N., and Mauntz, W. (2011).
\newblock The identification of model effective dimensions using global
  sensitivity analysis.
\newblock {\em Reliability Engineering and System Safety}, 96(4):440--449.

\bibitem[Kumar, 2008]{Kumar}
Kumar, A. (2008).
\newblock {\em Sequential calibration of computer models}.
\newblock PhD thesis, The Ohio State University.

\bibitem[Liu et~al., 2009]{Liu09}
Liu, F., Bayarri, M., and Berger, J. (2009).
\newblock Modularization in {B}ayesian analysis, with emphasis on analysis of
  computer models.
\newblock {\em Bayesian Analysis}, 4(1):119--150.

\bibitem[Loeppky et~al., 2006]{Lop06}
Loeppky, D., Bingham, D., and Welch, W. (2006).
\newblock Computer model calibration or tuning in practice.
\newblock Technical report, University of British Columbia.

\bibitem[Marrel et~al., 2009]{Marrel09}
Marrel, A., Iooss, B., Laurent, B., and Roustant, O. (2009).
\newblock Calculations of {S}obol indices for the {G}aussian process metamodel.
\newblock {\em Reliability Engineering and System Safety}, 94(3):742--751.

\bibitem[Pratola et~al., 2013]{Prat13}
Pratola, M., Sain, S., Bingham, D., Wiltberger, M., and Rigler, E. (2013).
\newblock Fast sequential computer model calibration of large nonstationary
  spatial-temporal processes.
\newblock {\em Technometrics}, 55(2):232--242.

\bibitem[Pronzato and M{ü}ller, 2012]{Pronz12}
Pronzato, L. and M{ü}ller, W. (2012).
\newblock Design of computer experiments: space filling and beyond.
\newblock {\em Statistics and Computing}, 22(3):681--701.

\bibitem[Rasmussen and Williams, 2006]{Rasm06}
Rasmussen, C. and Williams, C. (2006).
\newblock {\em {G}aussian Processes for Machine Learning}.
\newblock the MIT press.

\bibitem[Robert and Casella, 1998]{Robert+98}
Robert, C. and Casella, G. (1998).
\newblock {\em Monte Carlo Statistical Methods}.
\newblock Springer-Verlag.

\bibitem[Roy and Oberkampf, 2011]{Roy11}
Roy, C. and Oberkampf, W. (2011).
\newblock A comprehensive framework for verification, validation and
  uncertainty quantification in scientific computing.
\newblock {\em Computer Methods in Applied Mechanics and Engineering},
  200(25-28):2131--2144.

\bibitem[Sacks et~al., 1989]{Sacks89}
Sacks, J., W.J., W., Mitchell, T., and Wynn, H. (1989).
\newblock Design and analysis of computer experiments.
\newblock {\em Statistical Science}, 4(4):409--423.

\bibitem[Saltelli et~al., 2000]{Salt00}
Saltelli, A., Chan, K., and Scott, E. (2000).
\newblock {\em Sensitivity Analysis}.
\newblock Wiley, New York.

\bibitem[Santner et~al., 2003]{San03}
Santner, T., Williams, B., and Notz, W. (2003).
\newblock {\em The Design and Analysis of Computer Experiments}.
\newblock Springer-Verlag.

\bibitem[Schaback, 1995]{Scha95b}
Schaback, R. (1995).
\newblock Error estimates and condition numbers for radial basis function
  interpolation.
\newblock {\em Advances in Computational Mathematics}, 3(3):251--264.

\bibitem[Sheil and Muircheartaigh, 1977]{S7}
Sheil, J. and Muircheartaigh, I. (1977).
\newblock The distribution of non-negative quadratic forms in normal variables.
\newblock {\em Journal of the Royal Statistical Society, Series C},
  26(1):92--98.

\bibitem[Stein, 1999]{Stein99}
Stein, M. (1999).
\newblock {\em Interpolation of Spatial Data: Some Theory for Kriging}.
\newblock Springer, New York.

\bibitem[Vazquez and Bect, 2010]{Vasq10}
Vazquez, E. and Bect, J. (2010).
\newblock Convergence properties of the expected improvement algorithm with
  fixed mean and covariance functions.
\newblock {\em Journal of Statistical Planning and Inference},
  140(11):3088--3095.

\bibitem[Wong et~al., 2017]{Wong14}
Wong, R., Storlie, C., and Lee, T. (2017).
\newblock A frequentist approach to computer model calibration.
\newblock {\em Journal of the Royal Statistical Society: Series B},
  79:635--648.

\end{thebibliography}

\end{document}